\documentclass[envcountsame]{llncs}
\usepackage[latin1]{inputenc}   
\usepackage{graphicx}
\usepackage{latexsym}
\usepackage{amssymb}
\usepackage{amsmath}
\usepackage{caption2}
\usepackage{pstricks}
\usepackage{times}
\newcommand{\ignore}[1]{}

\usepackage{fancyhdr}
\author{Fredrik Kuivinen\thanks{Supported by the \emph{National Graduate School in Computer Science} (CUGS), Sweden.}}
\institute{Department of Computer and Information Science, Linköpings Universitet,\\S-581 83 Linköping, Sweden, \email{freku@ida.liu.se}}

\title{Approximability of Bounded Occurrence Max Ones} 

\ignore{
\SetKw{KwFrom}{from}
\SetKw{KwDown}{down}
\SetKw{KwGoto}{Goto}
\SetKw{Return}{Return}
\SetKwFor{For}{For}{do}{end}
\SetKwFor{ForEach}{Foreach}{do}{end}
\SetKwIF{If}{ElseIf}{Else}{If}{then}{Else if}{Else}{end}
\SetKwComment{comment}{$\rhd\ $}{}
\SetCommentSty{textnormal}
\dontprintsemicolon
}

\def\squareforqed{\hbox{\rlap{$\sqcap$}$\sqcup$}}
\def\qed{\ifmmode\squareforqed\else{\unskip\nobreak\hfil
\penalty50\hskip1em\null\nobreak\hfil\squareforqed
\parfillskip=0pt\finalhyphendemerits=0\endgraf}\fi}

\def\tup#1{\mathchoice{\mbox{\boldmath$\displaystyle#1$}}
{\mbox{\boldmath$\textstyle#1$}}
{\mbox{\boldmath$\scriptstyle#1$}}
{\mbox{\boldmath$\scriptscriptstyle#1$}}}

\renewcommand{\vec}[1]{\tup{#1}}
\usepackage{ifthen}

\newcommand{\prob}[1]{{\sc #1}}

\renewcommand{\mod}[1]{\ (\textrm{mod} \ #1)}
\newcommand{\opt}[0]{\textrm{{\sc opt}}}
\newcommand{\sol}[0]{\textrm{{\sc sol}}}

\newcommand{\cc}[1]{\textnormal{\textbf{#1}}} 
\newcommand{\proj}[2]{#1\big|_{#2}}
\newcommand{\Pol}[0]{\mathrm{Pol}}
\newcommand{\Inv}[0]{\mathrm{Inv}}

\newcommand{\PreserveBackslash}[1]{\let\temp=\\#1\let\\=\temp}

\begin{document}
\hyphenation{rep-re-sent-able}
\maketitle
\begin{abstract}
  We study the approximability of \prob{Max Ones} when the number of
  variable occurrences is bounded by a constant.  For conservative
  constraint languages (i.e., when the unary relations are included)
  we give a complete classification when the number of occurrences is
  three or more and a partial classification when the bound is two.

  For the non-conservative case we prove that it is either trivial or
  equivalent to the corresponding conservative problem under
  polynomial-time many-one reductions.


\end{abstract}

\section{Introduction}
Many combinatorial optimisation problems can be formulated as various
variants of constraint satisfaction problems (CSPs).  \prob{Max Ones}
is a boolean CSP where we are not only interested in finding a
solution but also the measure of the solution. In this paper we study
a variant of \prob{Max Ones} when the number occurrences of each
variable is bounded by a constant.

We denote the set of all $n$-tuples with elements from $\{0,1\}$ by
$\{0,1\}^n$. A subset $R \subseteq \{0,1\}^n$ is a \emph{relation} and
$n$ is the \emph{arity} of $R$. A \emph{constraint language} is a
finite set of relations. A constraint language is said to be
\emph{conservative} if every unary relation is included in the
language. In the boolean case this means that the relations $\{(0)\}$
and $\{(1)\}$ are in the language. The constraint satisfaction problem
over the constraint language $\Gamma$, denoted \prob{Csp$(\Gamma)$},
is defined to be the decision problem with instance $(V, C)$, where
$V$ is a set of variables and $C$ is a set of constraints $\{C_1,
\ldots, C_q\}$, in which each constraint $C_i$ is a pair $(R_i, s_i)$
with $s_i$ a list of variables of length $n_i$, called the constraint
scope, and $R_i$ an $n_i$-ary relation over the set $\{0, 1\}$,
belonging to $\Gamma$, called the constraint relation.
The question is whether there exists a solution to $(V, C)$ or not. A
solution to $(V, C)$ is a function $s : V \rightarrow \{0,1\}$ such
that, for each constraint $(R_i, (v_1, v_2, \ldots, v_{n_i})) \in
C$, the image of the constraint scope is a member of the constraint
relation, i.e., $(s(v_1), s(v_2), \ldots, s(v_{n_i})) \in R_i$.

The optimisation problem \prob{W-Max Ones} can be defined as follows:
\begin{definition}[\prob{W-Max Ones}] \label{def:wmaxones}
  \prob{W-Max Ones} over the constraint language $\Gamma$ is defined
  to be the optimisation problem with
  \begin{description}
  \item[Instance:] Tuple $(V, C, w)$, where $(V, C)$ is an instance of
    \prob{Csp$(\Gamma)$} and $w : V \rightarrow \mathbb{N}$ is a function.
  
  \item[Solution:] An assignment $f : V \rightarrow \{0,1\}$ to the
    variables which satisfies the \prob{Csp$(\Gamma)$} instance $(V,
    C)$.
    
  \item[Measure:] $\sum\limits_{v \in V} w(v) \cdot f(v)$
  \end{description}
\end{definition}
The function $w : V \rightarrow \mathbb{N}$ is called a \emph{weight
  function}. In the corresponding unweighted problem, denoted
\prob{Max Ones$(\Gamma)$}, the weight function is restricted to map
every variable to 1. The approximability of \prob{(W-)Max Ones} has
been completely classified by Khanna et al.~\cite{cspapprox}. Several
well-known optimisation problems can be rephrased as \prob{(W-)Max
  Ones} problems, in particular \prob{Independent Set}.  We will study
\prob{W-Max Ones$(\Gamma)$} with a bounded number of variable
occurrences, denoted by \prob{W-Max Ones$(\Gamma)$-$k$} for an integer
$k$. In this problem the instances are restricted to contain at most
$k$ occurrences of each variable. The corresponding bounded occurrence
variant of \prob{Csp$(\Gamma)$} will be denoted by
\prob{Csp$(\Gamma)$-$k$}.

Schaefer~\cite{gensat} classified the complexity of
\prob{Csp$(\Gamma)$} for every constraint language $\Gamma$. Depending
on $\Gamma$, Schaefer proved that \prob{Csp$(\Gamma)$} is either
solvable in polynomial time or is \cc{NP}-complete. The conservative
bounded occurrence variant of \prob{Csp$(\Gamma)$} has been studied by
a number of authors~\cite{delta-matroid,fanout-limitations,TF05,I97}.
One result of that research is that the difficult case to classify is
when the number of variable occurrences are restricted to two, in all
other cases the bounded occurrence problem is no easier than the
unrestricted problem. Kratochv\'{\i}l et al.~\cite{trivial-jump} have
studied \prob{$k$-Sat-$l$}, i.e., satisfiability where every clause
have length $k$ and there are at most $l$ occurrences of each
variable.  \prob{$k$-Sat-$l$} is a \emph{non-conservative} constraint
satisfaction problem. The complexity classification seems to be
significantly harder for such problems compared to the conservative
ones. In particular, Kratochv\'{\i}l et al~\cite{trivial-jump} proves
that there is a function $f$ such that \prob{$k$-Sat-$l$} is trivial
if $l \leq f(k)$ (every instance has a solution) and \cc{NP}-complete
if $l \geq f(k)+1$. Some bounds of $f$ is given
in~\cite{trivial-jump}, but the exact behaviour of $f$ is unknown.

\prob{Max Ones$(\Gamma)$-$k$} can represent many well-known problems.
For $k \geq 3$, we have for example, that \prob{Independent Set} in
graphs of maximum degree $k$ is precisely \prob{Max Ones$(\{\{(0,0),
  (1,0), (0,1)\}\})$-$k$}.  However, the more interesting case is
perhaps $k = 2$ due to its connection to matching problems.
(See~\cite{matching-ext} for definitions and more information about
the matching problems mentioned below.) Ordinary weighted maximum
matching in graphs is, for example, straightforward to formulate and
we get certain generalisations ``for free'' (because they can be
rephrased as ordinary matching problems), such as $f$-factors and
capacitated $b$-matchings. The general factor problem can also be
rephrased as a \prob{Max Ones$(\cdot)$-$2$} problem. A dichotomy
theorem for the existence problem of general factors has been proved
by Cornuéjols~\cite{gen-fact}. Some research has also been done on the
optimisation problem~\cite{half-rel}.

In this paper, we start the classification of bounded occurrence
\prob{Max Ones}. Our first result is a complete classification of
\prob{W-Max Ones$(\Gamma)$-$k$} when $k \geq 3$ and $\{(0)\}$ and
$\{(1)\}$ are included in $\Gamma$. We show that, depending on
$\Gamma$, this problem is either in \cc{PO}, \cc{APX}-complete or
\cc{poly-APX}-complete. Our second result is a partial classification
of \prob{W-Max Ones$(\Gamma)$-$2$}. We also give hardness results for
the non-conservative case.

The outline of the paper is as follows: in Section~\ref{sec:prel} we
define our notation and present the tools we use.
Section~\ref{sec:three-or-more} and~\ref{sec:two} contains our results
for three or more occurrences and two occurrences, respectively.
Section~\ref{sec:non-cons} contains our results for the general case,
i.e., when the constraint language is not necessarily conservative.
Section~\ref{sec:conc} contains some concluding remarks. Due to lack
of space most of the proofs can be found in the appendix.

\section{Preliminaries} \label{sec:prel}
For an integer $n$ we will use $[n]$ to denote the set $\{1, 2,
\ldots, n\}$. The Hamming distance between two vectors $\vec{x}$ and
$\vec{y}$ will be denoted by $d_H(\vec{x}, \vec{y})$. For a tuple or
vector $\vec{x}$ the $n$:th component will be denoted by $\vec{x}[n]$.

Unless explicitly stated otherwise we assume that the constraint
languages we are working with are \emph{conservative}, i.e., every
unary relation is a member of the constraint language (in the boolean
domain, which we are working with, this means that $\{(0)\}$ and
$\{(1)\}$ are in the constraint language).

We define the following relations
\begin{itemize}
\item $NAND^m = \{ (x_1, \ldots, x_m) \mid x_1 + \ldots + x_m < m \}$,
\item $EQ^m = \{ (x_1, \ldots, x_m) \mid x_1 = x_2 = \ldots = x_m \}$,
\item $IMPL = \{(0,0), (0,1), (1,1)\}$, $c_0 = \{(0)\}$, $c_1 = \{(1)\}$
\end{itemize}
and the function $h_n(x_1, x_2, \ldots, x_{n+1}) = \bigvee_{i=1}^{n+1}
(x_1 \land \ldots \land x_{i-1} \land x_{i+1} \land \ldots \land
x_{n+1})$. For a relation $R$ of arity $r$, we will sometimes use the
notation $R(x_1, \ldots, x_r)$ with the meaning $(x_1, \ldots, x_r)
\in R$, i.e., $R(x_1, \ldots, x_r) \iff (x_1, \ldots, x_r) \in R$. If
$r$ is the arity of $R$ and $I = \{i_1, \ldots, i_n\} \subseteq [r]$,
$i_1 < i_2 < \ldots < i_n$, then we denote the projection of $R$ to $I$
by $\proj{R}{I}$, i.e., $\proj{R}{I} =$ $\{ (x_{i_1}, x_{i_2},
  \ldots,$ $x_{i_n}) \mid$ $(x_1,$ $x_2, \ldots,$ $x_r) \in R \}$
  
  Representations (sometimes called implementations) have been central
  in the study of constraint satisfaction problems. We need a notion
  of representability which is a bit stronger that the usual one,
  because we have to be careful with how many occurrences we use of
  each variable.

\begin{definition}[$k$-representable]
  An $n$-ary relation $R$ is \emph{$k$-representable} by a set of
  relations $F$ if there is a collection of constraints $C_1, \ldots,
  C_l$ with constraint relations from $F$ over variables $\tup{x} =
  (x_1, x_2, \ldots, x_n)$ (called \emph{primary variables}) and
  $\tup{y} = (y_1, y_2, \ldots, y_m)$ (called \emph{auxiliary
    variables}) such that,
  \begin{itemize}
    \item the primary variables occur at most once in the constraints, 
    \item the auxiliary variables occur at most $k$ times in the
      constraints, and
    \item for every tuple $\tup{z}$, $\tup{z} \in R$ if and only if
      there is an assignment to $\tup{y}$ such that $\tup{x} =
      \tup{z}$ satisfies all of the constraints $C_1, C_2, \ldots,
      C_l$.
  \end{itemize}
\end{definition}
The intuition behind the definition is that if every relation in
$\Gamma_1$ is $k$-rep\-re\-sent\-able by relations in $\Gamma_2$ then
\prob{W-Max Ones$(\Gamma_2)$-$k$} is no easier than \prob{W-Max
  Ones$(\Gamma_1)$-$k$}. This is formalised in Lemma~\ref{lem:k-repr}.

\subsection{Approximability, Reductions, and Completeness}
A \emph{combinatorial optimisation problem} is defined over a set of
\emph{instances} (admissible input data) $\mathcal{I}$; each instance
$I \in \mathcal{I}$ has a finite set $\sol(I)$ of \emph{feasible
  solutions} associated with it.  The objective is, given an instance
$I$, to find a feasible solution of \emph{optimum} value with respect
to some measure function $m$ defined for pairs $(x,y)$ such that $x
\in \mathcal{I}$ and $y \in \sol(x)$. Every such pair is mapped to a
non-negative integer by $m$.
The optimal value is the largest one for
\emph{maximisation} problems and the smallest one for
\emph{minimisation} problems. A combinatorial optimisation problem is
said to be an \cc{NPO} problem if its instances and solutions can be
recognised in polynomial time, the solutions are polynomially-bounded in
the input size, and the objective function can be computed in
polynomial time (see, e.g.,~\cite{Kannetal99}).

\begin{definition}[$r$-approximate]
  A solution $s \in \sol(I)$ to an instance $I$ of an \cc{NPO} problem
  $\Pi$ is \emph{$r$-approximate} if $\max{\left \{ \frac{m(I,
        s)}{\opt(I)},\frac{\opt(I)}{m(I, s)} \right \} } \leq r$,
  where $\opt(I)$ is the optimal value for a solution to $I$.
\end{definition}
An approximation algorithm for an \cc{NPO} problem $\Pi$ has
\emph{performance ratio} $\mathcal{R}(n)$ if, given any instance $I$
of $\Pi$ with $|I|=n$, it outputs an $\mathcal{R}(n)$-approximate
solution.

\begin{definition}[\cc{PO}, \cc{APX}, \cc{poly-APX}]
  \cc{PO} is the class of \cc{NPO} problems that can be solved (to
  optimality) in polynomial time. An \cc{NPO} problem $\Pi$ is in the
  class \cc{APX} if there is a polynomial-time approximation algorithm
  for $\Pi$ whose performance ratio is bounded by a constant.
  Similarly, $\Pi$ is in the class \cc{poly-APX} if there is a
  polynomial-time approximation algorithm for $\Pi$ whose performance
  ratio is bounded by a polynomial in the size of the input.
\end{definition}

Completeness in \cc{APX} and \cc{poly-APX} is defined using
$AP$-reductions~\cite{Kannetal99}.  However, we do not need
$AP$-reductions in this paper, the simpler $L$- and $S$-reductions are
sufficient for us.

\begin{definition}[$L$-reduction]
  An \cc{NPO} problem $\Pi_1$ is said to be {\em $L$-reducible} to an
  \cc{NPO} problem $\Pi_2$, written $\Pi_1 \leq_L \Pi_2$, if two
  polynomial-time computable functions $F$ and $G$ and positive
  constants $\beta$ and $\gamma$ exist such that
  \begin{itemize}
  \item given any instance $I$ of $\Pi_1$, algorithm $F$ produces an instance
    $I'=F(I)$ of $\Pi_2$, such that $\opt(I') \leq \beta \cdot \opt(I)$.
    
  \item given $I'=F(I)$, and any solution $s'$ to $I'$, algorithm $G$
    produces a solution $s$ to $I$ such that $|m_1(I, s)-\opt(I)| \leq
    \gamma \cdot |m_2(I', s')-\opt(I')|$, where $m_1$ is the measure for
    $\Pi_1$ and $m_2$ is the measure for $\Pi_2$.
  \end{itemize}
\end{definition}
It is well-known (see, e.g., Lemma 8.2 in~\cite{Kannetal99}) that, if
$\Pi_1$ is $L$-reducible to $\Pi_2$ and $\Pi_1 \in \cc{APX}$ then
there is an $AP$-reduction from $\Pi_1$ to $\Pi_2$.

\emph{$S$-reductions} are similar to $L$-reductions but instead of the
condition $\opt(I') \leq \beta \cdot \opt(I)$ we require that
$\opt(I') = \opt(I)$ and instead of $|m_1(I, s) - \opt(I)| \leq \gamma
\cdot |m_2(I', s') - \opt(I')|$ we require that $m_1(I, s) = m_2(I',
s')$. If there is an $S$-reduction from $\Pi_1$ to $\Pi_2$ (written as
$\Pi_1 \leq_S \Pi_2$) then there is an $AP$-reduction from $\Pi_1$ to
$\Pi_2$.
An \cc{NPO} problem $\Pi$ is \emph{\cc{APX}-hard}
(\emph{\cc{poly-APX}-hard}) if every problem in \cc{APX}
(\cc{poly-APX}) is $AP$-reducible to it. If, in addition, $\Pi$ is in
\cc{APX} (\cc{poly-APX}), then $\Pi$ is called
\emph{\cc{APX}-complete} (\emph{\cc{poly-APX}-complete}).

We will do several reductions from \prob{Independent Set} (hereafter
denoted by \prob{MIS}) which is
\cc{poly-APX}-complete~\cite{syntactic-vs-comp}. We will also use the
fact that for any $k \geq 3$, \prob{MIS} restricted to graphs of
degree at most $k$ is \cc{APX}-complete~\cite{maxsnp}. We will denote
the latter problem by \prob{MIS-$k$}.

The following lemma shows the importance of $k$-representations in our
work.
\begin{lemma} \label{lem:k-repr}
  For constraint languages $\Gamma_1$ and $\Gamma_2$ if every relation
  in $\Gamma_1$ can be $k$-rep\-re\-sent\-ed by $\Gamma_2$ then \prob{W-Max
    Ones$(\Gamma_1)$-$k$} $\leq_S$ \prob{W-Max Ones$(\Gamma_2)$-$k$}.
\end{lemma}
\begin{proof}
  Given an arbitrary instance $I = (V, C, w)$ of \prob{W-Max
    Ones$(\Gamma_1)$-$k$}, we will construct an instance $I' = (V',
  C', w')$ of \prob{W-Max Ones$(\Gamma_2)$-$k$}, in polynomial time.
  For each $c \in C$, add the $k$-representation of $c$ to $C'$ and
  also add all variables which participate in the representation to
  $V'$ in such a way that the auxiliary variables used in the
  representation are distinct from all other variables in $V'$. Let
  $w'(x) = w(x)$ for all $x \in V$ and $w(x) = 0$ if $x \not \in V$
  (i.e., all auxiliary variables will have weight zero).
  
  It is not hard to see that: (a) every variable in $I'$ occurs at
  most $k$ times (b) 
  $\opt(I') = \opt(I)$, and (c) given a solution $s'$ to $I'$ we can
  easily construct a solution $s$ to $I$ (let $s(x) = s'(x)$ for every
  $x \in V$) such that $m(I, s) = m(I', s')$. Hence, there is an
  $S$-reduction from \prob{W-Max Ones$(\Gamma_1)$-$k$} to \prob{W-Max
    Ones$(\Gamma_2)$-$k$}. \qed
\end{proof}

\subsection{Co-clones and Polymorphisms}
Given an integer $k$, a function $f : \{0,1\}^k \rightarrow \{0, 1\}$
can be extended to a function over tuples as follows: let $\tup{t_1},
\tup{t_2}, \ldots, \tup{t_k}$ be $k$ tuples with $n$ elements each
then $f(\tup{t_1}, \tup{t_2}, \ldots, \tup{t_k})$ is defined to be the
tuple $(f(\tup{t_1}[1],$ $\tup{t_2}[1], \ldots,$ $\tup{t_k}[1]),
\ldots,$ $f(\tup{t_1}[n],$ $\tup{t_2}[n], \ldots,$ $\tup{t_k}[n]))$.
Given a $n$-ary relation $R$ we say that $R$ is \emph{invariant} (or,
closed) under $f$ if $\tup{t_1}, \tup{t_2}, \ldots, \tup{t_k} \in R
\Rightarrow f(\tup{t_1}, \tup{t_2}, \ldots, \tup{t_n}) \in R$.
Conversely, for a function $f$ and a relation $R$, $f$ is a
\emph{polymorphism} of $R$ if $R$ is invariant under $f$. For a
constraint language $\Gamma$ we say that $\Gamma$ is invariant under
$f$ if every relation in $\Gamma$ is invariant under $f$. We
analogously extend the notion of polymorphisms to constraint
languages, i.e., a function $f$ is a polymorphism of $\Gamma$ if
$\Gamma$ is invariant under $f$. Those concepts has been very useful
in the study of the complexity of various constraint satisfaction
problems (see, e.g.,~\cite{closure-prop}) and play an important role
in this work, too.

The set of polymorphisms for a constraint language $\Gamma$ will be
denoted by $\Pol(\Gamma)$, and for a set of functions $C$ the set of
all relations which are invariant under $C$ will be denoted by
$\Inv(B)$. The sets $\Pol(\Gamma)$ are \emph{clones} in the sense of
universal algebra. For a clone $C$, $\Inv(C)$ is called a relational
clone or a co-clone. Over the boolean domain Emil Post has classified
all such co-clones and their inclusion structure in~\cite{post}.

For a set of relations $\Gamma$ we define a closure operator $\langle
\Gamma \rangle$ as the set of relations that can be expressed with
relations from $\Gamma$ using existential quantification and
conjunction (note that we are only allowed to use the relations in
$\Gamma$, hence equality is not necessarily allowed). Intuitively
$\langle \Gamma \cup \{EQ^2\}\rangle$ is the set of relations which
can be simulated by $\Gamma$ in \prob{Csp$(\Gamma)$}.  An alternative
classification of this set is $\langle \Gamma \cup \{EQ^2\}\rangle =
\Inv(\Pol(\Gamma))$~\cite{PK79}. These few paragraphs barely scratch
the surface of the rich theory of clones and their relation to the
computational complexity of various constraint satisfaction problems,
for a more thorough introduction see~\cite{play1,play2,boolean-csp}.

We say that a set of relations $B$ is a \emph{plain basis} for a
constraint language $\Gamma$ if every relation in $\Gamma$ can be
expressed with relations from $B$ using relations from $B \cup \{=\}$
and conjunction. Note that this differs from the definition of the
closure operator $\langle \cdot \rangle$ as we do not allow
existential quantification. See \cite{pref-repr} for more information
on plain bases.

\begin{figure}[htb]
  \begin{minipage}[b]{0.5\linewidth}%
    \centering \includegraphics[width=6cm,height=6cm]{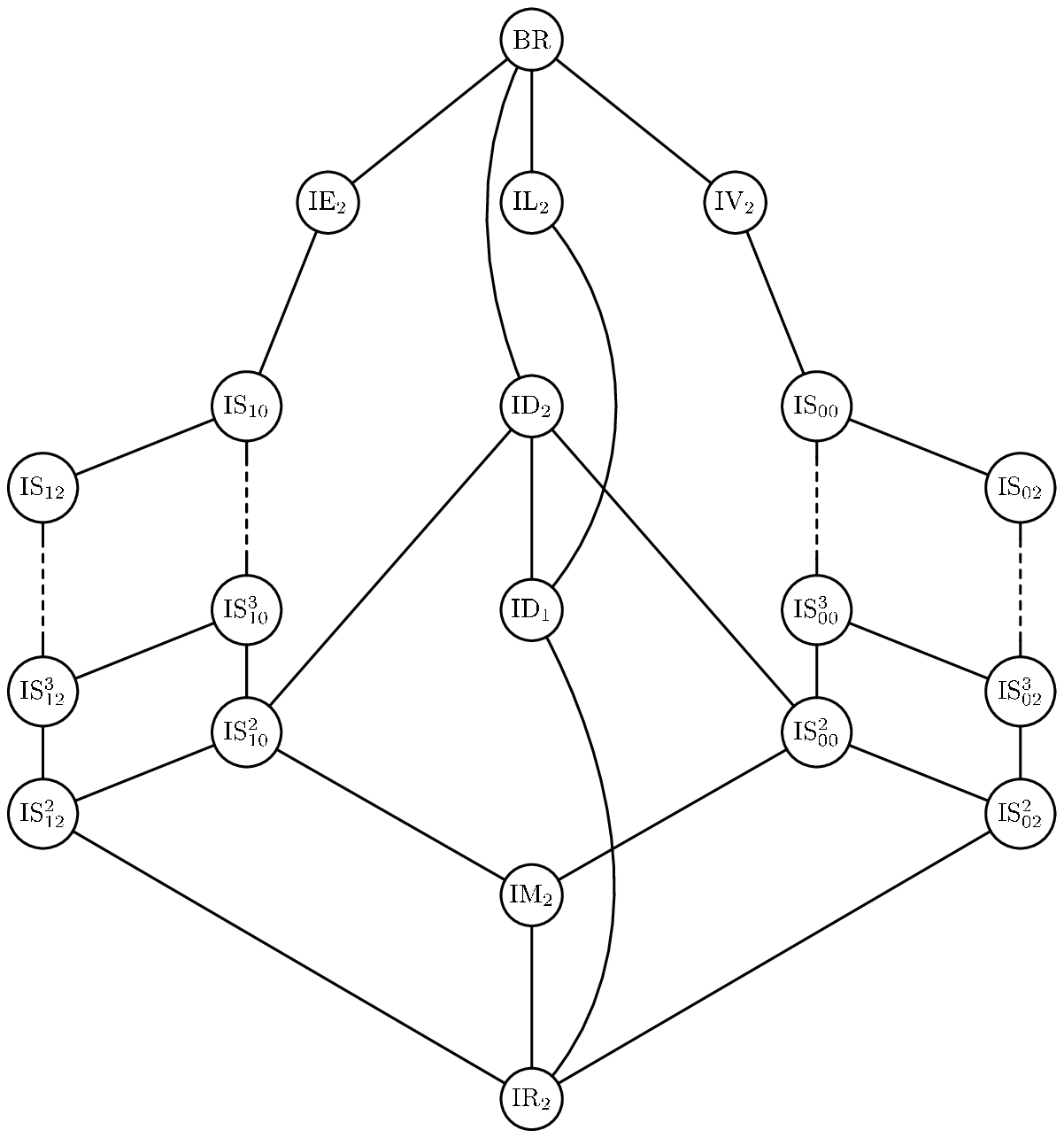}
    \caption{Lattice of idempotent co-clones} \label{fig:co-clones}
  \end{minipage}%
\begin{minipage}[b]{0.5\linewidth}%
    \centering \scalebox{0.85}{\begin{tabular}{l|p{2.5cm}|p{3cm}}
      Co-clone & Base for \ignore{corresponding} clone & Plain Basis \\
      \hline
      \hline
      
      $IE_2$       & $and$                           & \PreserveBackslash\raggedright $\{N_k \mid k \in \mathbb{N}\}$ $\cup$ $\{(\lnot x_1 \lor \ldots \lor \lnot x_k \lor y) \mid k \in \mathbb{N}\}$ \\
      $IS_{10}$    & $x \land (y \lor z)$            & \PreserveBackslash\raggedright $\{c_1, IMPL \}$ $\cup$ $\{N_k \mid k \in \mathbb{N}\}$                 \\
      $IS^m_{10}$  & $x \land (y \lor z), h_n$       & $\{c_1, IMPL, N_m \}^\ddagger$                 \\
      $IS_{12}$    & $x \land (y \lor \lnot z)$      & \PreserveBackslash\raggedright $\{EQ^2, c_1\}$ $\cup$ $\{N_k \mid k \in \mathbb{N}\}$ \\
      $IS^m_{12}$  & $x \land (y \lor \lnot z), h_n$ & $\{EQ^2, c_1, N_m \}^\ddagger$                 \\
      

      $IL_2$       & $x \oplus y \oplus z$                  & \PreserveBackslash\raggedright $\{x_1 \oplus \ldots \oplus x_k = c \mid k \in \mathbb{N}, c \in \{0,1\}\}$                 \\
      $ID_2$       & $xy \lor yz \lor xz$                   & \PreserveBackslash\raggedright $\{c_0,$ $c_1,$ $x \lor y,$ $IMPL,$ $NAND^2 \}$                 \\
      $ID_1$       & $xy \lor y (\lnot z) \lor y (\lnot z)$ & \PreserveBackslash\raggedright $\{c_0,$ $c_1,$ $x \oplus y = 0,$ $x \oplus y = 1\}$                 \\
      $IM_2$       & $and, or$                              & $\{c_0, c_1, IMPL \}$                 \\
      $IR_2$       & $or, x \land (y \oplus z \oplus 1)$    & $\{EQ^2, c_0, c_1\}$                 
    \end{tabular}}
    \captionof{table}{Plain bases for some relational clones. The list
      of plain bases are from~\cite{pref-repr}.$^\ddagger$\label{tab:bases}
}
  \end{minipage}
\end{figure}

We can not only study the co-clones when we try to classify \prob{Max
  Ones$(\Gamma)$-$k$} because the complexity of the problem do not
  only depend on the co-clone $\langle \Gamma \rangle$. However, the
  co-clone lattice with the corresponding plain bases and invariant
  functions will help us in our classification effort. Furthermore, as
  we mostly study the conservative constraint languages we can
  concentrate on the co-clones which contain $c_0$ and $c_1$.
  Figure~\ref{fig:co-clones} contains the conservative part of Post's
  lattice and Table~\ref{tab:bases} contains the plain bases for the
  relational clones which will be interesting to us (co-clones at and below
  $IV_2$ have been omitted as \prob{Max Ones} is in \cc{PO} there).

\vspace{-0.5cm}
\vspace{-0.5cm}

\section{Three or More Occurrences} \label{sec:three-or-more}
In this section we will prove a classification theorem for \prob{W-Max
  Ones$(\Gamma)$-$k$} where $k \geq 3$. The main result of this
section is the following theorem.

\footnotetext[5]{In~\cite{pref-repr} the listed plain basis for
  $IS^m_{12}$ is $\{EQ^2, c_1\} \cup \{N_k | k \leq m\}$ however, if
  we have $N_m$ then $N_{m-1}$ can be represented without auxiliary
  variables by $N_{m-1}(x_1, x_2, \ldots, x_{m-1}) \iff N_m(x_1, x_1,
  x_2, x_3, \ldots, x_{m-1})$, hence the set of relations listed in
  Table~\ref{tab:bases} is a plain basis for $IS_{12}^m$. The same
  modification has been done to $IS_{10}^m$.}

\begin{theorem} \label{th:classification}
Let $\Gamma$ be a conservative constraint language and $k \geq 3$,
\begin{enumerate}
\item If $\Gamma \subseteq IV_2$ then \prob{W-Max Ones$(\Gamma)$-$k$} is in
  \cc{PO}.
  
\item Else if $IS^2_{12} \subseteq \langle \Gamma \rangle \subseteq
  IS_{12}$ then \prob{(W-)Max Ones$(\Gamma)$-$k$} is \cc{APX}-complete
  if $EQ^2$ is not $k$-representable by $\Gamma$ and \prob{W-Max
    Ones$(\Gamma)$-$k$} is \cc{poly-APX}-complete otherwise.

\item Otherwise, \prob{W-Max Ones$(\Gamma)$} and \prob{W-Max
    Ones$(\Gamma)$-$k$} are equivalent under $S$-reductions.
\end{enumerate}
\end{theorem}
The first part of Theorem~\ref{th:classification} follows from Khanna
et al.'s results for \prob{Max Ones}~\cite{cspapprox}. Intuitively the
second part follows from the fact that \prob{W-Max Ones$(\{NAND^2\})$}
is equivalent to \prob{MIS}, hence if we have access to the equality
relation then the problem gets \cc{poly-APX}-complete. On the other
hand, if we do not have the equality relation then we essentially get
\prob{MIS-$k$}, for some $k$, which is \cc{APX}-complete. The third
part follows from Lemmas~\ref{lem:nand-or-neq}, \ref{lem:equal},
\ref{lem:eq-or-xnoty}, and~\ref{lem:strict}.


Dalmau and Ford proved the following lemma in~\cite{delta-matroid}.
\begin{lemma} \label{lem:nand-or-neq}
  If there is a relation $R$ in the constraint language $\Gamma$ such
  that $R \not \in IE_2$, then either $x \lor y$ or $x \neq y$ can be
  $3$-represented by $\Gamma$. By duality, if there is a relation $R
  \in \Gamma$ such that $R \not \in IV_2$, then either $NAND^2$ or $x
  \neq y$ can be $3$-represented.
\end{lemma}

We can use the lemma above to get a $3$-representation of either
$EQ^2$ or $IMPL$. We will later, in Lemma~\ref{lem:strict}, show that
those relations makes the problem as hard as the unbounded occurrence
variant.
\begin{lemma} \label{lem:equal}
  If there is a relation $R$ in the constraint language $\Gamma$ such
  that $R \not \in IE_2$ and $R \not \in IV_2$, then either $EQ^2$ or
  $IMPL$ can be $3$-represented by $\Gamma$.
\end{lemma}
\begin{proof}
  From Lemma~\ref{lem:nand-or-neq} we know that either $x \neq y$ or
  both $x \lor y$ and $NAND^2$ are $3$-representable. In the first
  case $\exists z: \ x \neq z \land z \neq y$ is a $3$-representation
  of $EQ^2$. In the second case $\exists z: \ NAND^2(x, z) \land (z
  \lor y)$ is a 3-representation of $IMPL(x, y)$.  \qed
\end{proof}

To get the desired hardness results for the $IS_{10}$ chain we need to
prove that we can represent $EQ_2$ or $IMPL$ in that case too. To this
end we have the following lemma.
\begin{lemma} \label{lem:eq-or-xnoty}
  If there is a relation $R$ in the constraint language $\Gamma$ such
  that $R \in IE_2$ and $R \not \in IS_{12}$, then either $EQ^2$ or
  $IMPL$ can be $3$-represented by $\Gamma$.
\end{lemma}
\begin{proof}
  Let $r$ be the arity of $R$ then, as $R \not \in IS_{12}$, there
  exists a set of minimal cardinality $I \subseteq [r]$, such that
  $\proj{R}{I} \not \in IS_{12}$.
  
  As $g(x, y) = x \land y$ is a base of the clone which corresponds to
  $IE_2$, $\proj{R}{I} \in IE_2$ implies that $g$ is a polymorphism of
  $\proj{R}{I}$.  Furthermore, as $f(x, y, z) = x \land (y \lor
  \lnot{z})$ is a base of the clone which corresponds to $IS_{12}$,
  $\proj{R}{I} \not \in IS_{12}$ implies that $f$ is \emph{not} a
  polymorphism of $\proj{R}{I}$. Hence, there exists tuples
  $\tup{t_1}, \tup{t_2}, \tup{t_3} \in \proj{R}{I}$ such that
  $f(\tup{t_1}, \tup{t_2}, \tup{t_3}) = \tup{t} \not \in \proj{R}{I}$.
  
  There exists a coordinate $l_1, 1 \leq l_1 \leq r$ such that
  $(\tup{t_1}[l_1], \tup{t_2}[l_1], \tup{t_3}[l_1]) = (1, 0, 1)$,
  because otherwise $f(\tup{t_1}, \tup{t_2}, \tup{t_3}) = \tup{t_1}$.
  Similarly there exists a coordinate $l_2, 1 \leq l_2 \leq r$ such
  that $(\tup{t_1}[l_2], \tup{t_2}[l_2], \tup{t_3}[l_2])$ is equal to
  one of $(0,1,0)$, $(0,1,1)$ or $(1,0,0)$.  Because otherwise
  $f(\tup{t_1}, \tup{t_2}, \tup{t_3}) = \tup{t_2}$.  From now on, the
  case $(\tup{t_1}[l_2], \tup{t_2}[l_2], \tup{t_3}[l_2]) = (1, 0, 0)$
  will be denoted by (*).  Finally, there also exists a coordinate
  $l_3, 1 \leq l_3 \leq r$ such that $(\tup{t_1}[l_3], \tup{t_2}[l_3],
  \tup{t_3}[l_3])$ is equal to one of $(0,0,1)$, $(0,1,1)$, $(1,0,0)$,
  $(1,0,1)$ or $(1,1,0)$, because otherwise $f(\tup{t_1}, \tup{t_2},
  \tup{t_3}) = \tup{t_3}$. The case $(\tup{t_1}[l_3], \tup{t_2}[l_3],
  \tup{t_3}[l_3]) = (1,0,0)$ will be denoted by (**).
  
  As $\proj{R}{I}$ is invariant under $g$ we can place additional
  restrictions on $l_1, l_2$ and $l_3$. In particular, there has to be
  coordinates $l_1, l_2$ and $l_3$ such that we have at least one of
  the cases (*) or (**), because otherwise $f(\tup{t_1}, \tup{t_2},
  \tup{t_3}) = g(\tup{t_1}, \tup{t_2})$, which is in $\proj{R}{I}$ and
  we have assumed that $f(\tup{t_1}, \tup{t_2}, \tup{t_3}) \not \in
  \proj{R}{I}$.  There is no problem in letting $l_2 = l_3$ since we
  will then get both (*) and (**). This will be assumed from now on.
  We can also assume, without loss of generality, that $l_1 = 1$ and
  $l_2 = l_3 = 2$. We can then construct a $3$-representation as
  $R_\phi(x, y) \iff \exists z_3 \ldots z_r: \proj{R}{I}(x, y, z_3, \ldots, z_r)
  \land c_{k_3}(z_3) \land c_{k_4}(z_4) \land \ldots \land
  c_{k_r}(z_r)$
  where $k_i = f(\tup{t_1}[i], \tup{t_2}[i], \tup{t_3}[i])$ for $3
  \leq i \leq r$. We will now prove that $R_\phi$ is equal to one of
  the relations we are looking for.
 
  If $(0, 1) \in R_\phi$, then we would have $\tup{t} \in
  \proj{R}{I}$, which is a contradiction, so $(0, 1) \not \in R_\phi$.
  We will now show that $(0, 0) \in R_\phi$. Assume that $(0, 0) \not
  \in R_\phi$. Then, $R^* = \proj{R}{I \setminus \{l_2\}}$ is not in
  $IS_{12}$ which contradicts the minimality of $I$. To see this
  consider the following table of possible tuples in $\proj{R}{I}$,
  \[
  \begin{array}{l|ccccc}
              & 1=l_1 & 2=l_2=l_3 & 3 & 4 & \ldots \\
    \hline
    \tup{t_1} & 1   & 1   & \tup{t_1}[3]                                & \tup{t_1}[4]                                & \ldots \\
    \tup{t_2} & 0   & 0   & \tup{t_2}[3]                                & \tup{t_2}[4]                                & \ldots \\
    \tup{t_3} & 1   & 0   & \tup{t_3}[3]                                & \tup{t_3}[4]                                & \ldots \\
    \tup{a}   & 0   & 1   & f(\tup{t_1}[3], \tup{t_2}[3], \tup{t_3}[3]) & f(\tup{t_1}[4], \tup{t_2}[4], \tup{t_3}[4]) & \ldots \\
    \tup{b}   & 0   & 0   & f(\tup{t_1}[3], \tup{t_2}[3], \tup{t_3}[3]) & f(\tup{t_1}[4], \tup{t_2}[4], \tup{t_3}[4]) & \ldots 
  \end{array}
  \]
  We know that $\tup{t_1}, \tup{t_2}, \tup{t_3} \in \proj{R}{I}$ and
  we also know that $\tup{a} \not \in \proj{R}{I}$. Furthermore, if
  $\tup{b} \not \in \proj{R}{I}$, then $\proj{f(\tup{t_1}, \tup{t_2},
    \tup{t_3})}{I \backslash \{l_2\}} \not \in R^*$ which means that
  $I$ is not minimal. The conclusion is that we must have $(0, 0) \in
  R_\phi$. In the same way it is possible to prove that unless $(1, 1)
  \in R_\phi$, $I$ is not minimal.

  To conclude, we have proved that $(0, 0), (1, 1) \in R_\phi$ and
  $(0, 1) \not \in R_\phi$, hence we either have $R_\phi = EQ^2$ or
  $R_\phi = \{(0, 0), (1, 0), (1, 1)\}$. \qed
\end{proof}

It is now time to use our implementations of $EQ^2$ or $IMPL$ to prove
hardness results. To this end we have the following lemma.
\begin{lemma} \label{lem:strict}
  If $EQ^2$ or $IMPL$ is $3$-representable by the constraint language
  $\Gamma$ then \prob{W-Max Ones$(\Gamma)$} $\leq_S$ \prob{W-Max
    Ones$(\Gamma)$-$3$}.
\end{lemma}
The proof can be found in the appendix. As either $EQ^2$ or $IMPL$ is
available we can construct a cycle of constraints among variables and
such a cycle force every variable in the cycle to obtain the same
value. Furthermore, each variable occurs only twice in such a cycle so
we have one occurrence left for each variable.

\section{Two Occurrences} \label{sec:two}
In this section, we study \prob{W-Max Ones$(\Gamma)$-$2$}. We are not
able to present a complete classification but a partial classification
is achieved.
We completely classify the co-clones $IL_2$ and $ID_2$.
For $\Gamma$ such that $\Gamma \not \subseteq
IL_2, ID_2$ we show that if there is a relation which is not a
$\Delta$-matroid relation (those are defined below) in $\Gamma$ then
\prob{W-Max Ones$(\Gamma)$-$2$} is \cc{APX}-hard if \prob{W-Max
  Ones$(\Gamma)$} is not tractable.

\subsection{Definitions and Results}
Most of the research done on \prob{Csp$(\Gamma)$-$k$} (e.g.,
in~\cite{fanout-limitations,delta-matroid,TF05}) has used the theory
of $\Delta$-matroids. Those objects are a generalisation of matroids
and has been widely studied, cf.~\cite{dmat-js,match-delta}. It turns
out that the complexity of \prob{W-Max Ones$(\Gamma)$-$2$} depend to a
large degree on if there is a relation which is not a $\Delta$-matroid
relation in the constraint language. $\Delta$-matroid relations are
defined as follows.
\begin{definition}[$\Delta$-matroid relation~\cite{delta-matroid}]
  Let $R \subseteq \{0, 1\}^r$ be a relation. If $\tup{x}, \tup{x'}
  \in \{0, 1\}^r$, then $\tup{x'}$ is a \emph{step from $\tup{x}$ to
    $\tup{y}$} if $d_H(\tup{x}, \tup{x'}) = 1$ and $d_H(\tup{x},
  \tup{x'}) + d_H(\tup{x'}, \tup{y}) = d_H(\tup{x}, \tup{y})$. $R$ is
  a \emph{$\Delta$-matroid relation} if it satisfies the following
  two-step axiom: $\forall \tup{x}, \tup{y} \in R$ and $\forall
  \tup{x'}$ a step from $\tup{x}$ to $\tup{y}$, either $\tup{x'} \in
  R$ or $\exists \tup{x''} \in R$ which is a step from $\tup{x'}$ to
  $\tup{y}$.
\end{definition}
As an example of a $\Delta$-matroid relation consider $NAND^3$. It is
not hard to see that $NAND^3$ satisfies the two-step axiom for every
pair of tuples as there is only one tuple which is absent from the
relation. $EQ^3$ is the simplest example of a relation which is not a
$\Delta$-matroid relation.
%
%
The main theorem of this section is the following partial
classification result for \prob{W-Max Ones$(\Gamma)$-2}. We say that a
constraint language $\Gamma$ \emph{is $\Delta$-matroid} if every
relation in $\Gamma$ is a $\Delta$-matroid relation.
\begin{theorem} \label{th:classification-2}
  Let $\Gamma$ be a conservative constraint language,
  \begin{enumerate}
  \item If $\Gamma \subseteq IV_2$ or $\Gamma \subseteq ID_1$ then
    \prob{W-Max Ones$(\Gamma)$-$2$} is in \cc{PO}.
  
  \item Else if $\Gamma \subseteq IL_2$ and,    
    \begin{itemize}
      \item $\Gamma$ is not $\Delta$-matroid then, \prob{W-Max Ones$(\Gamma)$-$2$} is \cc{APX}-complete.
      \item otherwise, \prob{W-Max Ones$(\Gamma)$-$2$} is in \cc{PO}.
    \end{itemize}

  \item Else if $\Gamma \subseteq ID_2$ and,
    \begin{itemize}
      \item $\Gamma$ is not $\Delta$-matroid then, \prob{W-Max Ones$(\Gamma)$-$2$} is \cc{poly-APX}-complete.
      \item otherwise, \prob{W-Max Ones$(\Gamma)$-2} is in \cc{PO}.
    \end{itemize}
  
    
  \item Else if $\Gamma \subseteq IE_2$ and $\Gamma$ is not
    $\Delta$-matroid then \prob{W-Max Ones$(\Gamma)$-$2$} is
    \cc{APX}-hard.
    
  \item Else if $\Gamma$ is not $\Delta$-matroid then it is
    \cc{NP}-hard to find feasible solutions to \prob{W-Max
      Ones$(\Gamma)$-$2$}.
  \end{enumerate}
\end{theorem}
Part~1 of the theorem follows from the known results for \prob{W-Max
  Ones}~\cite{Kannetal99}. Part~4 follows from results for
\prob{Csp$(\Gamma)$-$2$}~\cite[Theorem~4]{fanout-limitations}. The
other parts follows from the results in Sections~\ref{sec:id-il}
and~\ref{sec:is} below.


\subsection{Tractability Results for \prob{W-Max Ones$(\Gamma)$-$2$}} \label{sec:edmonds}

Edmonds and Johnson~\cite{well-solved} has shown that the following
integer linear programming problem is solvable in polynomial time:
maximise $\vec{w} \vec{x}$ subject to the constraints $\vec{0} \leq
\vec{x} \leq \vec{1}$, $\vec{b_1} \leq A \vec{x} \leq \vec{b_2}$ and
$\vec{x}$ is an integer vector.
%
Here $A$ is a matrix with integer entries such that the sum of the
absolute values of each column is at most $2$. $\vec{b_1}$,
$\vec{b_2}$ and $\vec{w}$ are arbitrary real vectors of appropriate
dimensions. We will denote this problem by \prob{ILP-2}. With the
polynomial solvability of \prob{ILP-2} it is possible to prove the
tractability of a number of \prob{W-Max Ones$(\Gamma)$-$2$} problems.

\subsection{Classification of $ID_2$ and $IL_2$} \label{sec:id-il}
When $\Pol(\Gamma) = \Pol(ID_2)$ or $\Pol(\Gamma) = \Pol(IL_2)$ we
prove a complete classification result. We start with the hardness
results for $ID_2$, which consists of the following lemma.

\begin{lemma} \label{lem:id2-hard}
  Let $\Gamma$ be a constraint language such that $\Pol(\Gamma) =
  \Pol(ID_2)$. If there is a relation $R \in \Gamma$ which is not a
  $\Delta$-matroid relation, then \prob{W-Max Ones$(\Gamma)$-$2$} is
  \cc{poly-APX}-complete.
\end{lemma}
The main observations used to prove the lemma is that since
$\Pol(\Gamma) = \Pol(ID_2)$ we can 2-represent every
two-literal clause. This has been proved by Feder
in~\cite{fanout-limitations}.  Furthermore, if we have access to every
two-literal clause and also have a non-$\Delta$-matroid relation then
it is possible to make variables participate in three clauses, which
was also proved in~\cite{fanout-limitations}. The hardness result then
follows with a reduction from \prob{MIS}.

We will use some additional notation in the following proofs. For a
tuple $\tup{x} = (x_1, x_2, \ldots, x_k)$ and a set of coordinates $A
\subseteq [k]$, $\tup{x} \oplus A$ is defined to be the tuple $(y_1,
y_2, \ldots, y_k)$ where $y_i = x_i$ if $i \not \in A$ and $y_i =
1-x_i$ otherwise.  We extend this notation to relations: if $R
\subseteq \{0,1\}^n$ and $A \subseteq [n]$ then $R \oplus A =
\{\tup{t} \oplus A \mid \tup{t} \in R\}$.
  
We will now define a constraint language denoted by $\mathcal{Q}$.  We
will later prove that \prob{W-Max Ones$(\mathcal{Q})$-$2$} is in
\cc{PO}. $\mathcal{Q}$ is the smallest constraint language such that:
  \begin{itemize}
  \item $\emptyset$, $c_0$, $c_1$, $EQ^2$ and $\{(0,1), (1,0)\}$ are
    in $\mathcal{Q}$.
    
  \item Every relation definable as $\{\tup{t} \mid d_H(\tup{0},
    \tup{t}) \leq 1\}$ is in $\mathcal{Q}$.

  \item If $R, R' \in \mathcal{Q}$ then their cartesian product $\{ (\tup{t},
    \tup{t'}) \mid \tup{t} \in R, \tup{t'} \in R'\}$ is also in $\mathcal{Q}$.
    
  \item If $R \in \mathcal{Q}$ and $n$ is the arity of $R$ then $R
    \oplus A \in \mathcal{Q}$ for every $A \subseteq [n]$.
    
  \item If $R \in \mathcal{Q}$, $n$ is the arity of $R$ and $f : [n] \rightarrow
    [n]$ is a permutation on $[n]$ then $\{ (t_{f(1)}, t_{f(2)},
    \ldots, t_{f(n)}) \mid \tup{t} \in R\}$ is in $\mathcal{Q}$.
  \end{itemize}
  
  The relation between $\mathcal{Q}$ and the $\Delta$-matroid
  relations in $ID_2$ is given by the following lemma.
\begin{lemma} \label{lem:Q}
  If $R \in ID_2$ is a $\Delta$-matroid relation, then $R \in
  \mathcal{Q}$.
\end{lemma}

As for the tractability part we have the following lemma.
\begin{lemma} \label{lem:id2-po}
  Let $\Gamma$ be a constraint language such that $\Gamma \subseteq
  ID_2$, if all relations in $\Gamma$ are $\Delta$-matroid relations
  then \prob{W-Max Ones$(\Gamma)$-$2$} is in \cc{PO}.
\end{lemma}
The idea behind the proof is that \prob{W-Max Ones$(\mathcal{Q})$-$2$}
can be seen as an \prob{ILP-2} problem and is therefore solvable in
polynomial time.

As for $IL_2$ the result is the same, non $\Delta$-matroids give rise
to \cc{APX}-complete problems and absence of such relations makes the
problem tractable. Also in this case the tractability follows from a
reduction to \prob{ILP-2}.


\subsection{$IE_2$, $IS_{12}$ and $IS_{10}$} \label{sec:is}
The structure of the $\Delta$-matroids do not seem to be as simple in
$IS_{12}$ and $IS_{10}$ as they are in $ID_2$ and $IL_2$. There exists
relations in $IS_{12}$ which are $\Delta$-matroid relations but for
which we do not know of any polynomial time algorithm. One such
relation is $R(x, y, z, w) \iff NAND^3(y,z,w) \land NAND^3(x,z,w)
\land NAND^2(x, y)$.  However, we get tractability results for some
relations with the algorithm for \prob{ILP-2}. In particular if the
constraint language is a subset of $\{NAND^m \mid m \in \mathbb{N} \}
\cup \{IMPL\}$ then \prob{W-Max Ones$(\cdot)$-$2$} is in \cc{PO}.

We manage to prove hardness results for every non-$\Delta$-matroid
relation contained in those co-clones.  The main part of our hardness
results for the non-$\Delta$-matroid relations is the following lemma.
\begin{lemma} \label{lem:abc5}
%
  Let $R(x_1, x_2, x_3) \iff NAND^2(x_1, x_2) \land NAND^2(x_2, x_3)$,
  then \mbox{\prob{W-Max Ones$(\{c_0, c_1, R\})$-$2$}} is \cc{APX}-complete.
\end{lemma}
Note that $R$ is not a $\Delta$-matroid relation. With
Lemma~\ref{lem:abc5} and a careful enumeration of the types of
non-$\Delta$-matroid relations that exists in $IE_2$, we can deduce
the desired result: if there is a non-$\Delta$-matroid relation in the
constraint language, then \prob{W-Max Ones$(\cdot)$-$2$} is
\cc{APX}-hard. The proof builds upon the work
in~\cite{fanout-limitations,3-dim-matching,mis-3}.

\section{Non-conservative Constraint Languages} \label{sec:non-cons}
In this section we will take a look at the non-conservative case,
i.e., we will look at constraint languages which do not necessarily
contain $c_0$ and $c_1$.
A relation $R$ is said to be \emph{1-valid} if it contains the all
ones tuple, i.e., $R$ is 1-valid if $(1,1,\ldots,1) \in R$. A
constraint language is said to be 1-valid if every relation in the
language is 1-valid.

\begin{theorem} \label{th:non-cons}
  For any constraint language $\Gamma$ which is not 1-valid, if
  \prob{W-Max Ones$(\Gamma \cup \{c_0, c_1\})$-$k$} is \cc{NP}-hard
  for some integer $k$ then so is \prob{W-Max Ones$(\Gamma)$-$k$}.
\end{theorem}
Note that for constraint languages $\Gamma$ which are 1-valid
\prob{W-Max Ones$(\Gamma)$} is trivial: the all-ones solution is
optimal. The idea in the proof is that we can simulate $c_1$
constraints by giving the variable a large weight. Furthermore, if
there are relations which are not $1$-valid then we can represent
$c_0$ constraints when we have access to $c_1$ constraints. It fairly
easy to see why this fails to give us any inapproximability results:
due to the large weight used to simulate $c_1$ any feasible solution
is a good approximate solution.

\section{Conclusions} \label{sec:conc}
We have started the study of the approximability properties of bounded
occurrence \prob{Max Ones}. We have presented a complete
classification for the weighted conservative case when three or more
variable occurrences are allowed. Furthermore, a partial
classification of the two occurrence case has been presented. In the
latter case we have proved that non-$\Delta$-matroid relations give
rise to problems which are \cc{APX}-hard if the unbounded occurrence
variant is not tractable. We have also given complete classifications
for the $IL_2$ and $ID_2$ co-clones.

There are still lots of open questions in this area. For example, what
happens with the complexity if the weights are removed? Many
constraint satisfaction problems such as \prob{Max Ones} and \prob{Max
  Csp} do not get any harder when weights are added. Such results are
usually proved by scaling and replicating variables and constraints a
suitable number of times.  However, such techniques do not work in the
bounded occurrence setting and we do not know of any substitute which
is equally general.

Except for the $IS_{12}$ and $IS_{10}$ chains the open questions in
the two occurrence case are certain constraint languages $\Gamma$ such
that $\Gamma$ only contains $\Delta$-matroid relations and
$\Pol(\Gamma) = \Pol(BR)$. It would be very interesting to find out
the complexity of \prob{W-Max Ones$(\cdot)$-$2$} for some of the
classes of $\Delta$-matroid relations which have been proved to be
tractable for \prob{Csp$(\cdot)$-$2$}
in~\cite{fanout-limitations,delta-matroid,I97,TF05}. Instead of trying
to classify the entire $IS_{12}$ or $IS_{10}$ chain one could start
with $IS_{12}^3$ or $IS_{10}^3$. The approximability of the
non-conservative case is also mostly open.  In light
of~\cite{trivial-jump} the computational structure of those problems
seems to be quite complex.

\clearpage

\section*{Appendix}




\subsection*{Proofs for Results in Section~\ref{sec:three-or-more}}

\begin{proof}[Of Lemma~\ref{lem:strict}]
  Let $I = (V, C, w)$ be an instance of \prob{W-Max Ones$(\Gamma)$}.
  We will start with the case when $IMPL$ is $3$-representable.
  
  If $IMPL$ is $3$-representable we can reduce $I$ to an instance $I'
  = (V', C', w')$ of \prob{W-Max Ones$(\Gamma)$-$2$} as follows: for
  each variable $v_i \in V$, let $o_i$ be the number of occurrences of
  $v_i$ in $I$, we introduce the variables $v_i^1, \ldots, v_i^{o_i}$
  in $V'$. We let $w'(v_i^1) = w(v_i)$ and $w'(v_i^j) = 0$ for $j \neq
  1$. We also introduce the constraints $IMPL(v_i^k, v_i^{k+1})$ for
  $k, 1 \leq k \leq o_i-1$ and $IMPL(v_i^{o_i}, v_i^{1})$ into $C'$.
  For every $i, 1 \leq i \leq |V|$ those constraints makes the
  variables $v_i^1, \ldots, v_i^{o_i}$ have the same value in every
  feasible solution of $I'$.
  
  For every constraint $c = (R, s) \in C$ the constraint
  scope $s = (v_{l_1}, \ldots, v_{l_m})$ is replaced by $s' =
  (v_{l_1}^{k_1}, \ldots, v_{l_m}^{k_m})$ and $(R, s')$ is added
  to $C'$. The numbers $k_1, \ldots, k_m$ are chosen in such a way
  that every variable in $V'$ occur exactly three times in $I'$.  This
  is possible since there are $o_i$ variables in $V'$ for every
  $v_i \in V$.
  
  It is clear that the procedure described above is an $S$-reduction
  from \prob{W-Max Ones$(\Gamma)$} to \prob{W-Max Ones$(\Gamma)$-$3$}.
  
  $I$ can easily be $S$-reduced to an instance $I'$ of \prob{W-Max
    Ones($\Gamma \cup \{EQ^2\}$)-$3$}. And as $EQ^2$ is
  $3$-representable by $\Gamma$ we are done, as every constraint
  involving $EQ^2$ can be replaced by the $3$-representation of $EQ^2$
  and any auxiliary variables used in the representation can be
  assigned the weight zero.  \qed
\end{proof}

We need a couple of lemmas before we can state the proof of the
classification theorem (Theorem~\ref{th:classification}). The
following lemma will be used in several places to prove hardness
results.

\begin{lemma} \label{lem:is-implements-nand}
  Let $\Gamma$ be a constraint language such that $\Pol(\Gamma) =
  \Pol(IS_{1\alpha}^m)$ for some integer $m$ and $\alpha \in \{0,2\}$,
  then $NAND^m$ can be $2$-represented by $\Gamma$.
\end{lemma}
\begin{proof}
  As $\Pol(\Gamma) = \Pol(IS_{1\alpha}^m)$, $\Gamma$ is invariant
  under $h_m$ and not invariant under $h_{m-1}$. Let $r$ be the arity
  of $R$ and let $X \subseteq [r]$ be a set of minimal cardinality
  such that there exist tuples $\tup{x_1}, \tup{x_2}, \ldots,
  \tup{x_m} \in \proj{R}{X}$ which satisfies $h_{m-1}(\tup{x_1},
  \tup{x_2}, \ldots, \tup{x_m}) = \tup{z} \not \in \proj{R}{X}$. If
  there is a coordinate $i \in X$ such that $\tup{x_1}[i] =
  \tup{x_2}[i] = \ldots = \tup{x_m}[i]$ then $\tup{z}[i] =
  \tup{x_1}[i]$ and as $X$ is minimal we must have $\tup{z} \oplus i
  \in \proj{R}{X}$. However, this means that $h_m(\tup{x_1},
  \tup{x_2}, \ldots, \tup{x_m}, \tup{z} \oplus i) = \tup{z} \not \in
  \proj{R}{X}$ which is a contradiction with the assumption that $R$
  is invariant under $h_m$. We conclude that no coordinate is
  constant in every $\tup{x_1}, \tup{x_2}, \ldots, \tup{x_m}$.
  
  Now assume that there is a coordinate $j \in X$ such that
  $\tup{z}[j] = 0$, then for $X$ to be minimal we must have $\tup{z}
  \oplus j \in \proj{R}{X}$. However, $h_m(\tup{x_1}, \tup{x_2},
  \ldots, \tup{x_m}, \tup{z} \oplus j) = \tup{z} \not \in
  \proj{R}{X}$, a contradiction, hence there is no $j \in X$ such that
  $\tup{z}[j] = 0$.
  
  We can assume that $|X| \geq m$ because every relation of arity less
  than $m$ which is invariant under $h_m$ is also invariant under
  $h_{m-1}$~\cite[Proposition~3.6]{simple-bases}.
  
  We do now know three things, no coordinate in $X$ is constant in
  every $\tup{x_1}, \tup{x_2}, \ldots, \tup{x_m}$, $\tup{z} =
  (1,1,\ldots,1)$ and $|X| \geq m$. As $\tup{z} = (1,1,\ldots,1)$
  there is at most one zero for every given coordinate $i \in X$
  among $\tup{x_1}[i], \tup{x_2}[i], \ldots, \tup{x_m}[i]$, however as
  there is no constant coordinate and $|X| \geq m$ we must have at
  least one zero in every $\tup{x_1}, \tup{x_2}, \ldots, \tup{x_m}$.
  We can in fact assume that there is exactly one zero entry, because
  if it is two distinct coordinates $i,j \in X$ such that
  $\tup{x_1}[i] = \tup{x_1}[j] = 0$ then as $\tup{z} = (1,1,\ldots,1)$
  no other tuple can have $\tup{x_k}[i] = 0$ or $\tup{x_k}[j] = 0$.
  The conclusion is that $\proj{R}{X \setminus \{j\}}$ is not invariant
  under $h_{m-1}$ either.

  This implies that $\tup{x_i} = (1,1,\ldots,1) \oplus i$. It is not
  hard to see that by using the fact that $R$ is invariant under
  \emph{and} we can get any tuple $\tup{y} = (y_1, y_2, \ldots, y_m)$
  such that $y_1+y_2+\ldots+y_m < m$ by applying \emph{and} to the
  $\tup{x_i}$s an appropriate number of times.  Hence, we must have
  $\proj{R}{X} = NAND^m$. \qed
\end{proof}

\begin{lemma} \label{lem:nand-impl}
  If $\Pol(\{R\}) = \Pol(IS_{12}^m)$ for some $m \geq 2$ and $R$
  cannot represent $EQ^2$, then $\langle \{R, c_0, c_1\} \rangle =
  \langle \{NAND^m, c_1\} \rangle$.
\end{lemma}
\begin{proof}
  We will denote $NAND^m$ by $N$. Let $r$ be the arity of $R$ then $B
  = \{N, EQ^2, c_1\}$ is a plain basis for $IS_{12}^m$ (see
  Table~\ref{tab:bases}). As $B$ is a plain basis for $R$ there is an
  implementation of R on the following form,
\begin{align}
R(x_1, \ldots, x_r) \iff
&N(x_{k_1^1}, x_{k_1^2}, \ldots, x_{k_1^m}) \land \ldots \land N(x_{k_n^1}, \ldots, x_{k_n^m}) \land \notag \\
&EQ^2(x_{l_1^1}, x_{l_1^2}) \land \ldots \land EQ^2(x_{l_c^1}, x_{l_c^2}) \notag \\
&c_1(x_{c_{1}}) \land \ldots \land c_1(x_{c_w}) \notag
\end{align}
for some $n$, $c$ and $w$ such that $k_i^j \in [r]$, $l_i^j \in [r]$ and $c_i \in [r]$.

Assume that the representation above is minimal in the sense that it
contains a minimal number of constraints. Hence, the only equalities
that are possible are of the form $EQ^2(x_i, x_j)$ for $i \neq j$. If
there is such an equality there are a number of cases to consider,
\begin{enumerate}
\item $\proj{R}{\{i, j\}} = \{(0, 0), (1, 1)\}$,
\item $\proj{R}{\{i, j\}} = \{(1, 1)\}$, and
\item $\proj{R}{\{i, j\}} = \{(0, 0)\}$.
\end{enumerate}
We cannot have equalities of type 1 because then $EQ^2$ would be
representable by $R$. Furthermore, equalities of type 2 and 3 can be
replaced by constraints of the form $c_1(x_i) \land c_1(x_j)$ and
$N(x_i, \ldots, x_i) \land N(x_j, \ldots, x_j)$, respectively.

The conclusion is that $R$ can be represented without $EQ^2$ and hence
it is representable by $\{N, c_1\}$ alone. We have thus proved that
$\langle \{R, c_0, c_1\} \rangle \subseteq \langle \{N, c_1\}
\rangle$. The other inclusion, $\langle \{N, c_1\} \rangle \subseteq
\langle \{R, c_0, c_1\} \rangle$, is given by
Lemma~\ref{lem:is-implements-nand}.  \qed
\end{proof}

As for the containment we have the following lemma.
\begin{lemma} \label{lem:alg}
  Let $\Gamma$ be a constraint language if $\Gamma \subseteq
  IS^m_{12}$ for some $m$ and $\Gamma$ cannot represent $EQ^2$
  then \prob{W-Max Ones$(\Gamma)$-$k$} is in \cc{APX}.
\end{lemma}
\begin{proof}
  Lemma~\ref{lem:nand-impl} tells us that $\langle \Gamma \rangle =
  \langle \{ NAND^m, c_1 \} \rangle$, hence an instance $J$ of
  \prob{W-Max Ones$(\Gamma)$-$k$} can be reduced to an instance $J'$ of
  \prob{W-Max Ones$(\{NAND^m, c_1\})$-$k'$} for some constant $k'$. To
  prove the lemma it is therefore sufficient show that \prob{W-Max
    Ones$(\{NAND^m, c_1\})$-$l$} is in \cc{APX} for every fixed $l$.
  
  Let $I = (V, C, w)$ be an arbitrary instance of \prob{W-Max
    Ones$(\{NAND^m, c_1\})$-$l$}, for some $l$, and assume that $V =
  \{x_1, \ldots, x_n\}$.  By Schaefer's result~\cite{gensat} we can
  decide in polynomial time whether $I$ have a solution or not. Hence,
  we can safely assume that $I$ has a solution.  If a variable occurs
  in a constant constraint, say $c_1(x)$, then $x$ must have the same
  value in every model of $I$. Thus, we can eliminate all such
  variables and assume that $I$ only contains constraints of the type
  $NAND^m(x_1, \ldots, x_m)$.
  
  We will give a polynomial-time algorithm that creates a satisfying
  assignment $s$ to $I$ with measure at least $\frac{1}{l+1}\opt(I)$.
  Hence we have a $\frac{1}{l+1}$-approximate algorithm proving that
  \prob{W-Max Ones$({IS_{12}})$-$l$} is in \cc{APX}.
  
  The algorithm is as follows: Repeatedly delete from $I$ any variable
  $x_i$ having maximum weight and all variables that appear together
  with $x_i$ in a clause of size two. In $s$ we assign $1$ to $x_i$
  and $0$ to all variables appearing together with $x_i$ in a clause
  of size two.
  
  For simplicity, assume that the algorithm chooses variables $x_1,
  x_2, \ldots, x_t$ before it stops.  If the algorithm at some stage
  choose a variable $x$ with weight $w(x)$, then, in the worst case,
  it is forced to set $l$ (remember that no variable occurs more than
  $l$ times in $I$) variables to $0$ and each of these variables have
  weight $w(x)$. This implies that $(l+1) \cdot \sum_{i=1}^t
  w(x_i) \geq \sum_{i=1}^n w(x_i)$ and
  \[
  m(I, s) = \sum_{i=1}^t w(x_i) \geq \frac{1}{l+1} \sum_{i=1}^n w(x_i) \geq \frac{\opt(I)}{l+1}.
  \]
  \qed
\end{proof}

\begin{lemma} \label{lem:nand-hard}
  Let $\Gamma$ be a constraint language such that $IS_{12}^2 \subseteq
  \langle \Gamma \rangle \subseteq IS_{12}$ then \prob{W-Max
    Ones$(\Gamma)$-$k$} is \cc{APX}-hard for $k \geq 3$.
\end{lemma}
\begin{proof}
  Note that \prob{MIS-3} is exactly the same as \prob{Max
    Ones$(\{NAND^2\})$-$3$}. The lemma then follows from the fact that
  \prob{MIS-3} is \cc{APX}-hard, Lemma~\ref{lem:is-implements-nand},
  and Lemma~\ref{lem:k-repr}. \qed
\end{proof}

We are now ready to give the proof of the classification theorem for
three or more occurrences.

\begin{proof}[Of Theorem~\ref{th:classification}, part 1]
  Follows directly from Khanna et al's results for \prob{Max
    Ones}~\cite{cspapprox}.  \qed
\end{proof}
\begin{proof}[Of Theorem~\ref{th:classification}, part 2]
  The \cc{APX}-hardness follows from Lemma~\ref{lem:nand-hard}.
  Containment in \cc{APX} follows from Lemma~\ref{lem:alg}.  If $EQ^2$
  is $k$-representable by $\Gamma$ then the result follows from
  Lemma~\ref{lem:strict} and Khanna et al's results for \prob{Max
    Ones}~\cite{cspapprox}.  \qed
\end{proof}
\begin{proof}[Of Theorem~\ref{th:classification}, part 3]
  There are two possibilities, the first one is that $\Gamma \not
  \subseteq IE_2$ and $\Gamma \not \subseteq IV_2$, the second case
  is that $\Gamma \subseteq IE_2$ and $\Gamma \not \subseteq IS_{12}$.
  
  In the first case we can use the $3$-representation of $EQ^2$ or
  $IMPL$ from Lemma~\ref{lem:equal}. The result then follows from
  Lemma~\ref{lem:strict}. In the second case the result follows from
  Lemma~\ref{lem:eq-or-xnoty} and Lemma~\ref{lem:strict}.  \qed
\end{proof}

\subsection*{Proofs for Results in Section~\ref{sec:two}}

We will start with the case when $\Pol(\Gamma) = \Pol(ID_2)$. We need
the following lemma before we can give the proof of
Lemma~\ref{lem:id2-hard}.

\begin{lemma} \label{lem:2-clause}
  Let $\Gamma$ be a constraint language such that $\Pol(\Gamma)
  = \Pol(ID_2)$ then $x \lor y$, $IMPL$ and $NAND^2$ are
  $2$-representable by $\Gamma$.
\end{lemma}
\begin{proof}
  A part of the proof of Theorem~3 in~\cite{fanout-limitations} is the
  following: let $F$ be a constraint language such that there are
  relations $R_1, R_2, R_3 \in F$ with the following properties:
  \begin{itemize}
    \item $R_1$ is not closed under $f(x, y) = x \lor y$.
    \item $R_2$ is not closed under $g(x, y) = x \land y$.
    \item $R_3$ is not closed under $h(x, y, z) = x+y+z \mod{2}$.
  \end{itemize}
  then $F$ can $2$-represent every two-literal clause.  As we have
  assumed that $\Pol(\Gamma) = \Pol(ID_2)$ there are relations
  in $\Gamma$ which full fills the conditions above. The lemma
  follows.  \qed
\end{proof}

\begin{proof}[Of Lemma~\ref{lem:id2-hard}]
  We will do an $S$-reduction from the \cc{poly-APX}-complete problem
  \prob{MIS}, which is precisely \prob{Max Ones$(\{NAND^2\})$}. Let $I
  = (V, C)$ be an arbitrary instance of \prob{Max Ones$(\{NAND^2\})$}.
  We will construct an instance $I' = (V', C', w)$ of \prob{W-Max
    Ones$(\Gamma)$-$2$}. From Lemma~\ref{lem:2-clause} we know that we
  can $2$-represent every two-literal clause. It is easy to modify $I$
  so that each variable occur at most three times. For a variable $x
  \in V$ which occur $k$ times, introduce $k$ fresh variables $y_1,
  y_2, \ldots, y_k$ and add the constraints $IMPL(y_1, y_2),$
  $IMPL(y_2, y_3),$ $\ldots,$ $IMPL(y_k, y_1)$. Each occurrence of $x$
  is then replaced with one of the $y_i$ variables.  In every solution
  each of the $y_i$ variables will obtain the same value, furthermore
  they occur three times each.  Hence, if we can create a construction
  which allows us to let a variable participate in three clauses we
  are done with our reduction.
  
  In Theorem~4 in~\cite{fanout-limitations} it is shown that given a
  relation which is not a $\Delta$-matroid we can make variables
  participate in three clauses if we have access to all clauses.
  
  If we assign appropriate weights to the variables in $V'$ it is
  clear that $\opt(I) = \opt(I')$ and each solution to $I'$
  corresponds to a solution of $I$ with the same measure. Hence, we
  get an $S$-reduction. \qed
\end{proof}

We will now give the proof of Lemma~\ref{lem:Q} which describes the
structure of the $\Delta$-matroid relations in $ID_2$.

For a relation $R \in \mathcal{Q}$ if $R$ can be decomposed (possibly
after a permutation of the coordinates of $R$) into a cartesian
product of other relations, $P_1, P_2, \ldots, P_n \in \mathcal{Q}$
then $P_1, P_2, \ldots, P_n$ will be called the \emph{factors} of $R$.

\begin{proof}[Of Lemma~\ref{lem:Q}]
  In this proof we will denote the majority function by $m$, i.e.,
  $m(x,y,z) = (x \land y) \lor (y \land z) \lor (x \land z)$. Note
  that every relation in $ID_2$ is invariant under $m$. Let $R$ be a
  relation which contradicts the lemma, i.e., $R \in ID_2$, $R$ is a
  $\Delta$-matroid and $R \not \in \mathcal{Q}$. Let $n$ be the arity
  of $R$.  We can assume without loss of generality that $R$ consists
  of one factor, i.e., it is not possible to decompose $R$ into a
  cartesian product of other relations. In particular, $R$ do not
  contain any coordinate which has the same value in all tuples.
  
  As every relation of arity less than or equal to two is in
  $\mathcal{Q}$ we can assume that $n \geq 3$. If for every pair of
  tuples $\tup{t}, \tup{t'} \in R$ we have $d_H(\tup{t}, \tup{t'})
  \leq 2$ then $R \in \mathcal{Q}$ which is a contradiction. To see
  this let $\tup{t_1}, \tup{t_2}, \tup{t_3}$ be three distinct tuples
  in $R$ (if there are less than three tuples in $R$ then either $R$
  is not a $\Delta$-matroid relation or there is some coordinate which
  is constant in all tuples). Then $\tup{t_2} = \tup{t_1} \oplus A$,
  $\tup{t_3} = \tup{t_1} \oplus B$ for some $A, B \subseteq [n]$ such
  that $|A|, |B| \leq 2$ and $|A \cap B| \leq 1$. If $|A \cup B| \leq
  2$ for all such sets then $R$ is either of arity 2 or there is a
  coordinate in $R$ which is constant. Hence, assume that $|A \cup B|
  = 3$, which implies $|A|=|B|=2$. Let $\tup{t} = m(\tup{t_1},
  \tup{t_2}, \tup{t_3})$. We will prove that for every tuple $\tup{t'}
  \in R$ we have $d_H(\tup{t'}, \tup{t}) \leq 1$. To this end, let
  $\tup{t'} = \tup{t} \oplus C$, with $|C| = 2$ ($|C| \leq 1$ implies
  $d_H(\tup{t'}, \tup{t}) \leq 1$), be an arbitrary tuple in $R$. If
  $|A \cap C| = 0$ (or, $|B \cap C| = 0$) then $d_H(\tup{t_2},
  \tup{t'}) \geq 3$ ($d_H(\tup{t_3}, \tup{t'}) \geq 3$). Hence, we
  must have $|A \cap C|, |B \cap C| \geq 1$ but this implies
  $d_H(\tup{t}, \tup{t'}) \leq 1$ or $d_H(\tup{t}, \tup{t'}) \geq 3$, but
  the latter is not possible.  We conclude that for every tuple
  $\tup{t'} \in R$ we have $d_H(\tup{t}, \tup{t'}) \leq 1$, hence $R
  \in \mathcal{Q}$ which contradicts our assumption that $R \not \in
  \mathcal{Q}$.
  
  Hence, there exists tuples $\tup{t}, \tup{t'} \in R$ such that
  $d_H(\tup{t}, \tup{t'}) \geq 3$. If for every pair of such tuples it
  is the case that every step, $\tup{s}$, from $\tup{t}$ to $\tup{t'}$
  we have $\tup{s} \in R$, then as no coordinate is constant, we must
  have $(0,0,\ldots,0) \in R$ and $(1,1,\ldots,1) \in R$. However, if
  $(0,0,\ldots,0), (1,1,\ldots,1) \in R$ and every step from the
  former to the latter is in $R$ then every tuple with one coordinate
  set to $1$ is in $R$, too. We can continue this way and get every
  tuple with two coordinates set to one and then every tuple with $k$
  coordinates set to $1$ for $k \in [n]$. Hence, we must have $R =
  \{0,1\}^n \in \mathcal{Q}$.
  
  We can therefore assume that there exists an coordinate $l$ such
  that the step $\tup{s} = \tup{t} \oplus l$ from $\tup{t}$ to
  $\tup{t'}$ is not in $R$. Then, as $R$ is a $\Delta$-matroid
  relation, there exist another coordinate $K$ such that $\tup{s}
  \oplus \{K\} \in R$ is a step from $\tup{s}$ to $\tup{t'}$. Let $X$
  denote the set of coordinates $i$ such that $\tup{t} \oplus i
  \not \in R$ but $\tup{t} \oplus \{K, i\} \in R$, furthermore choose
  $\tup{t}$ and $K$ such that $|X|$ is maximised and let $X' = X \cup
  \{K\}$.
  
  Our goal in the rest of the proof is to show that if $X' = [n]$ then
  $R \in \mathcal{Q}$ and otherwise it is possible to decompose $R$
  into a cartesian product with $\proj{R}{X'}$ in one factor and
  $\proj{R}{[n] \setminus X'}$ in the other factor. As we have assumed
  that $R$ cannot be decomposed into a cartesian product we get a
  contradiction and hence the relation $R$ cannot exist.

  \subsection*{Case 1: $|X'| = 2$}
  We will start with the case when $|X'| = 2$. Assume, without loss of
  generality, that $X' = \{x, K\}$ then $\tup{t}, \tup{t} \oplus \{x,
  K\} \in R$ and $\tup{t} \oplus x \not \in R$. We will now prove that
  we cannot have any tuples $\tup{v}$ in $R$ such that
  $\proj{\tup{v}}{X'} = \proj{(\tup{t} \oplus x)}{X'}$. If we had such
  a tuple then $m(\tup{v}, \tup{t}, \tup{t} \oplus \{x, K\}) = \tup{w}
  \in R$ due to the fact that $R \in ID_2$ and $m$ is a polymorphism
  of $ID_2$. Furthermore, $\tup{w}$ must have the same value as
  $\tup{t}$ on every coordinate except for possibly $x$ and $K$, this
  follows from the fact that $\tup{t}$ has the same value as $\tup{t}
  \oplus \{x, K\}$ on every coordinate except for $x$ and $K$. Hence,
  the only coordinates for which we do not know the value of $\tup{w}$
  are $x$ and $K$.  However, $\tup{v}[K] = \tup{t}[K]$ (due do the
  construction of $\tup{v}$ and the fact that $K \in X'$). Hence we
  must get $\tup{w}[K] = \tup{t}[K]$. For $\tup{w}[x]$ note that
  $\tup{v}[x] = (\tup{t} \oplus \{x, K\})[x]$, hence $\tup{w}[x] =
  (\tup{t} \oplus x)[x]$. We can finally conclude $\tup{w} = \tup{t}
  \oplus x$ which is a contradiction with the construction of $X'$.
  
  Similar arguments as the above will be used repeatedly in this
  proof. However, the presentation will not be as detailed as the one
  above.
  
  We split the remaining part of case 1 into two subcases, when
  $\tup{t} \oplus K \not \in R$ (subcase 1a) and $\tup{t} \oplus K \in
  R$ (subcase 1b).

  \subsubsection*{Subcase 1a: $\tup{t} \oplus K \not \in R$}
  Assume that $\tup{t} \oplus K \not \in R$, then $\proj{(\tup{t}
    \oplus K)}{X'} \not \in \proj{R}{X'}$, because given a tuple
  $\tup{v}$ such that $\proj{\tup{v}}{X'} = \proj{(\tup{t} \oplus
    K)}{X'}$ then $m(\tup{t}, \tup{t} \oplus \{x, K\}, \tup{v}) =
  \tup{t} \oplus K$, which is not in $R$ by the assumption we made.
  
  Furthermore, for any tuple $\tup{v} \in R$, $\tup{v} \oplus x$ is a
  step from $\tup{v}$ to either $\tup{t}$ or $\tup{t} \oplus \{x,
  K\}$, but $\tup{v} \oplus x \not \in R$ (because either
  $\proj{\tup{v}}{X'} = \proj{\tup{t}}{X'}$ which would imply $\tup{v}
  \oplus x \not \in R$, or $\proj{\tup{v}}{X'} = \proj{(\tup{t} \oplus
    \{x, K\})}{X'}$ which implies $\proj{(\tup{v} \oplus x)}{X'} =
  \proj{(\tup{t} \oplus K)}{X'} \not \in \proj{R}{X'}$).
  
  The only way to get from $\tup{v} \oplus x$ to something which is in
  $R$ is by flipping coordinate $K$, hence $\tup{v} \oplus \{x, K\}
  \in R$.  This is the end of the case when $\tup{t} \oplus K \not \in
  R$, because what we have proved above is that $R$ can be decomposed
  into a cartesian product with the coordinates $X'$ in one factor and
  $[n] \setminus X'$ in the other factor.

  \subsubsection*{Subcase 1b: $\tup{t} \oplus K \in R$}
  We know that $\proj{(\tup{t} \oplus x)}{X'} \not \in \proj{R}{X'}$.
  We will now show that for any $\tup{v} \in R$ such that,
  $\proj{\tup{v}}{X'}$ is either $\proj{\tup{t}}{X'}$ or $\proj{(\tup{t}
    \oplus \{x, K\})}{X'}$, we have $\tup{v} \oplus \{x, K\} \in R$.
  
  To this end, let $\tup{v}$ be an arbitrary tuple in $R$ satisfying
  one of the conditions above. We will consider the two possible cases
  separately.

  \begin{itemize}
  \item If $\proj{\tup{v}}{X'} = \proj{\tup{t}}{X'}$ then $\tup{v}
    \oplus x$ is a step from $\tup{v}$ to $\tup{t} \oplus \{x, K\}$
    and $\tup{v} \oplus x \not \in R$. Furthermore, the only way to
    get into $R$ is by flipping $K$ hence $\tup{v} \oplus \{x,K\} \in
    R$.
    
  \item If $\proj{\tup{v}}{X'} = \proj{(\tup{t} \oplus \{x, K\})}{X'}$
    then $\tup{v} \oplus K$ is a step from $\tup{v}$ to $\tup{t}$ and
    $\tup{v} \oplus K \not \in R$. Furthermore, the only way to get
    into $R$ is by flipping $x$ hence $\tup{v} \oplus \{x,K\} \in R$.
  \end{itemize}
  
  Now, let $\tup{v}$ be an arbitrary tuple in $R$ such that
  $\proj{\tup{v}}{X'} = \proj{(\tup{t} \oplus K)}{X'}$ then $\tup{v}
  \oplus K$ is a step from $\tup{v}$ to $\tup{t}$. If $\tup{v} \oplus
  K \in R$ or $\tup{v} \oplus x \in R$ then we are done with this
  step, so assume that $\tup{v} \oplus K, \tup{v} \oplus x \not \in
  R$. However, as $R$ is a $\Delta$-matroid relation there has to
  exist a coordinate $l$ such that $\tup{v} \oplus \{K,l\} \in R$.
  Then we get, $\proj{(\tup{v} \oplus \{K,l\})}{X'} =
  \proj{\tup{t}}{X'}$ which implies $\tup{v} \oplus \{x, l\} \in R$ by
  the argument above. However, this means that $|X|$ is not maximal we
  could have chosen $\tup{v}$, $l$ and $X'$ instead of $\tup{t}$, $K$
  and $X$. We conclude that $\tup{v} \oplus K \in R$.
  
  Finally, let $\tup{v}$ be an arbitrary tuple in $R$ such that
  $\proj{\tup{v}}{X'} = \proj{\tup{t}}{X'}$ then $\tup{v} \oplus K \in
  R$. To see this note that $m(\tup{t} \oplus K, \tup{v}, \tup{v}
  \oplus \{x, K\}) = \tup{v} \oplus K$.
  
  We have now proved that $R$ can be decomposed into a cartesian
  product with the coordinates $X'$ in one factor and $[n] \setminus
  X'$ in the other factor for this case too.
  
  As we have assumed that the arity of $R$ is strictly greater than
  two we have $X' \neq [n]$. Hence, $[n] \setminus X' \neq \emptyset$.

  \subsection*{Case 2: $|X'| > 2$}
  The rest of the proof will deal with the case when $|X'| > 2$. We
  will begin with establishing a number of claims of $R$. Assuming
  that $X' \neq [n]$, our main goal is still to show that $R$ can be
  decomposed into a cartesian product with $X'$ in one factor and $[n]
  \setminus X'$ in one factor. If $X' = [n]$ we will show that $R \in
  \mathcal{Q}$.
 
  \textbf{Claim 1: if $d_H(\proj{\tup{x}}{X}, \proj{\tup{t}}{X}) = 1$
    and $\tup{x}[K] = \tup{t}[K]$ then $\tup{x} \not \in R$}
  
  Let $\tup{x}$ be a tuple which satisfies the precondition in the
  claim, assume that $\tup{x} \in R$, and let $i \in X$ be a
  coordinate where $\tup{x}$ differs from $\tup{t}$. By the
  construction of $X$ we have that $\tup{t} \oplus \{K, i\} \in R$,
  hence we get $m(\tup{t}, \tup{t} \oplus \{K, i\}, \tup{x}) = \tup{t}
  \oplus \{i\} \in R$, which is a contradiction.
  
  \textbf{Claim 2: if $d_H(\proj{\tup{x}}{X}, \proj{\tup{t}}{X}) = m$,
    for any $m$ such that $2 \leq m \leq |X|$, then $\tup{x} \not \in
    R$}
  
  We will prove this claim by induction on $m$. For the base case,
  let $m=2$. Let $x \in X$ be some coordinate such that $\tup{x}[x]
  \neq \tup{t}[x]$, if $\tup{x} \in R$ and $\tup{x}[K] = \tup{t}[K]$
  then $m(\tup{t}, \tup{x}, \tup{t} \oplus \{x, K\}) = \tup{t} \oplus
  x \not \in R$.  Hence $\tup{x}[K] = \tup{t}[K]$ is not possible.
  
  On the other hand if $\tup{x}[K] \neq \tup{t}[K]$ then $\tup{x}
  \oplus K$ is a step from $\tup{x}$ to $\tup{t}$. By the argument in
  the preceding paragraph we get $\tup{x} \oplus K \not \in R$ (note
  that $K \not \in X$ hence we have $d_H(\proj{(\tup{x} \oplus K)}{X},
  \proj{\tup{t}}{X}) = m$). Furthermore as $R$ is a $\Delta$-matroid
  we can flip some coordinate $l \in X$ such that $\tup{t}[l] \neq
  \tup{x}[l]$ to get a tuple which is in $R$ ($l \not \in X$ will not
  work as the argument in the preceding paragraph still applies in
  that case). However, $d_H(\proj{(\tup{x} \oplus \{K,l\})}{X},
  \proj{\tup{t}}{X}) = 1$ hence by claim~1 we get a contradiction.
  
  Now, assume that claim 2 holds for $m = m'$. We will prove that it
  also holds for $m = m'+1$ such that $2 < m \leq |X|$. Note that we
  can use exactly the same argument as above except for the very last
  sentence in which we appeal to claim 2 with $m = m'$ instead of
  using claim 1.  As we have assumed that claim 2 holds for $m = m'$
  we are done.

  \textbf{Claim 3: there is a tuple $\tup{z} \in \proj{R}{X'}$ such
    that for any tuple $\tup{x} \in \proj{R}{X'}$ we have
    $d_H(\tup{z}, \tup{x}) \leq 1$}
  
  If $|X'| > 2$, then there are tuples $\tup{t} \oplus \{i, K\}$ and
  $\tup{t} \oplus \{j, K\}$ for distinct $i,j,K \in X'$ in $R$. Hence,
  the tuple $\tup{z'} = m(\tup{t}, \tup{t} \oplus \{i,K\}, \tup{t}
  \oplus \{j,K\}) = \tup{t} \oplus K \in R$. Let $\tup{z} =
  \proj{\tup{z'}}{X'}$. We will now show that $d_H(\tup{z},
  \proj{\tup{x}}{X'}) \leq 1$ for every tuple $\tup{x}$ in $R$. To
  this end, let $\tup{x}$ be an arbitrary tuple in $R$. By claim~2 we
  must have $d_H(\proj{\tup{x}}{X}, \proj{\tup{t}}{X}) \leq 1$,
  furthermore if $\tup{x}[K] = \tup{z'}[K] \neq \tup{t}[K]$ then we
  are done as $d_H(\proj{\tup{x}}{X'}, \proj{\tup{z'}}{X'}) = 1$ in
  this case. On the other hand, if $\tup{x}[K] = \tup{t}[K]$ then
  claim~1 and claim~2 tells us that we must have
  $d_H(\proj{\tup{x}}{X}, \proj{\tup{t}}{X}) = 0$ in which case
  claim~3 follows.
  
  \textbf{Claim 4: if $\tup{x} \in R$ and $\proj{\tup{x}}{X'} =
    \tup{z} \oplus \{i\}$ for some $i \in X'$, then $\tup{x} \oplus
    \{i, j\} \in R$ for every $j \in X'$.}
  
  Given $j \in X', j \neq i$, there is at least one tuple $\tup{v} \in
  R$ such that $\tup{v}[j] \neq \tup{x}[j]$ since otherwise the
  coordinate $j$ would be constant and $R$ could be decomposed into a
  cartesian product. Hence, $\tup{x'} = \tup{x} \oplus j$ is a step
  from $\tup{x}$ to $\tup{v}$, but claim~3 tells us that $\tup{x'}
  \not \in R$ and the only way to full fill the two-step axiom is if
  $\tup{x} \oplus \{i,j\} \in R$ (due to claim~3 we cannot have
  $d_H(\tup{x}, \tup{v}) = 1$).
  
  We will now prove that $R$ can be decomposed into cartesian product
  where the coordinates $X'$ make up one factor and $[n] \setminus
  X'$ make up the other factor. Let $P = \proj{R}{X'}$.  Our goal
  is to show that for any $\tup{p} \in P$ and $\tup{v} \in
  \proj{R}{[n] \setminus X'}$ we have $(\tup{p}, \tup{v}) \in R$ (we
  have assumed that $X' = \{1,2,3, \ldots, |X'|\}$ here).
  
  To this end, let $\tup{v}$ and $\tup{v'}$ be arbitrary tuples in $R$.
  By claim~3 there either is a coordinate $i \in X'$ such that
  $\proj{(\tup{v} \oplus i)}{X'} = \tup{z}$ or $\proj{\tup{v}}{X'} =
  \tup{z}$.  The same is true for $\tup{v'}$; either there is an
  coordinate $i' \in X'$ such that $\proj{(\tup{v'} \oplus i')}{X'}
  = \tup{z}$ or $\proj{\tup{v'}}{X'} = \tup{z}$,
  
  If $\proj{\tup{v'}}{X'} = \proj{\tup{v}}{X'}$ or
  $\proj{\tup{v'}}{[n] \setminus X'} = \proj{\tup{v'}}{[n] \setminus
    X'}$ then we are done, so assume that neither holds.
  
  If $\proj{\tup{v'}}{X'} \neq \tup{z}$ and $\proj{\tup{v}}{X'} \neq
  \tup{z}$ then $\tup{s} = \tup{v} \oplus i'$ is a step from
  $\tup{v}$ to $\tup{v'}$ but by claim~3 $\tup{s} \not \in R$ and the
  only way to go a step from $\tup{s}$ to $\tup{v'}$ and get into $R$
  is $\tup{s'} = \tup{s} \oplus i$, hence $\tup{s'} \in R$.
  
  For the other case, when $\proj{\tup{v'}}{X'} = \tup{z}$ and
  $\proj{\tup{v}}{X'} \neq \tup{z}$, if there is a coordinate $j \in
  X'$ such that $\tup{v'} \oplus j \in R$ then we are back to the
  previous case, so assume that such a $j$ do not exist. As $R$ is a
  $\Delta$-matroid there must be a coordinate $x \not \in X'$ such
  that $\tup{s} = \tup{v'} \oplus \{i, x\} \in R$, because for some
  appropriate $x$, $\tup{s}$ is a step from $\tup{v'} \oplus i$ to
  $\tup{v}$. Due to claim~4 we will then have $\tup{v'} \oplus \{x,
  y\} \in R$ for every $y \in X'$. However, this contradicts the
  maximality of $X$ since we could have chosen $\tup{v'}$, $x$, and
  $X'$ instead of $\tup{t}$, $K$, and $X$. The conclusion is that if
  $X' \neq [n]$, then $R$ can be decomposed into a cartesian product.
  On the other hand, if $X' = [n]$, then we can easily deduce from
  claim 3 that $R \in \mathcal{Q}$.  \qed
\end{proof}

\begin{proof}[Of Lemma~\ref{lem:id2-po}]
  Let $I$ be an arbitrary instance of \prob{W-Max Ones$(\Gamma)$-$2$}.
  We will show that the problem is in \cc{PO} by reducing it to an
  instance $I'$ of \prob{ILP-2}. For any relation $R \in \Gamma$ of
  arity $n$ we know, from Lemma~\ref{lem:Q}, that $R \in \mathcal{Q}$.
  We can assume that $R$ is not the cartesian product of any other two
  relations, because if it is then every use of $R$ can be substituted
  by the factors in the cartesian product. If $R$ is unary we can
  replace $R(x)$ by $x = 0$ or $x = 1$. If $R = EQ^2$ then we can
  replace $R(x, y)$ by $x = y$ and if $R(x, y) \iff x \neq y$ then we
  replace $R(x,y)$ by $x = y - 1$.
  
  Now, assume that none of the cases above occur. We will show that
  \begin{align}
  R(t_1, t_2, \ldots, t_n) \iff \sum_{i=1}^n a_i t_i \leq b  \label{eq:R}
  \end{align}
  for some $a_i \in \{-1, 1\}$ and integer $b$. Let $N$ be set of
  negated coordinates of $R$, i.e., let $N \subseteq [n]$ such that
  \[
  R = \{(f(t_1, 1), f(t_2, 2), \ldots, f(t_n, n)) \mid d_H(\tup{0}, \tup{t}) \leq 1\}
  \]
  where $f : \{0,1\} \times [n] \rightarrow \{0, 1\}$ and $f(x, i) =
  \lnot x$ if $i \in N$ and $f(x, i) = x$ otherwise. According to the
  definition of $\mathcal{Q}$, $R$ can be written on this form. Let $a_i = -1$
  if $i \in N$ and $a_i = 1$ otherwise. Furthermore, let $b = 1 -
  |N|$.  It is now easy to verify that~\eqref{eq:R} holds.
  
  As every variable occur at most twice in $I$ every variable occur at
  most twice in $I'$ too. Furthermore, the coefficient in front of any
  variable in $I'$ is either $-1$, $0$ or $1$, hence the sum of the
  absolute values in any column in $I'$ is bounded by $2$. $I'$ is
  therefore an instance of \prob{ILP-2}. If we let the weight function
  of $I'$ be the same as the weight function in $I$ it easily seen
  that any solution $s$ to $I'$ is also a solution to $I$ with the
  same measure. \qed
\end{proof}

As we are done with $ID_2$ we will continue with $IL_2$. A linear
system of equations over GF(2) with $n$ equations and $m$ variables
can be represented by a matrix $A$, a constant column vector $\vec{b}$
and a column vector $\vec{x} = (x_1, \ldots, x_m)$ of variables. The
system of equations is then given by $A\vec{x} = \vec{b}$. Assuming
that the rows of $A$ are linearly independent the set of solutions to
$A\vec{x} = \vec{b}$ are
\[
\left\{ (\vec{x}', \vec{x}'') \mid \vec{x''} \in \mathbb{Z}_2^{n-m}
\textrm{ and } \vec{x}' = A'\vec{x}'' + \vec{b}' \right\}.
\]
where $\vec{x'} = (x_1, \ldots, x_m)$, $\vec{x}'' = (x_{m+1}, \ldots,
x_n)$ and $A'$ and $\vec{b}'$ are suitably chosen.

If there is a column in $A'$ with more than one entry which is equal
to $1$ (or, equivalently more than one non-zero entry), then we say
that the system of equations is \emph{coupled}.

\begin{lemma} \label{lem:il2-2}
  Let $\Gamma$ be a conservative constraint language such that $\Gamma
  \subseteq IL_2$.  If there is a relation $R \in \Gamma$ such that
  $R$ is the set of solutions to a coupled linear system of equations
  over $GF(2)$ then \prob{W-Max Ones$(\Gamma)$} $\leq_L$ \prob{W-Max
    Ones$(\Gamma)$-$2$}, otherwise \prob{W-Max Ones$(\Gamma)$-$2$} is
  in \cc{PO}.
\end{lemma}
\begin{proof}
  First note that every relation $R \in IL_2$ is the set of solutions
  to a linear system of equations over GF(2)~\cite{pref-repr}.
  
  We will start with the hardness proof. To this end we will construct
  a $2$-representation of $EQ^3$. Let $R \in \Gamma$ be defined by
  \[
  R = \left\{ (\vec{x}', \vec{x}'') \mid \vec{x''} \in \mathbb{Z}_2^{n-m}
    \textrm{ and } \vec{x}' = A'\vec{x}'' + \vec{b}' \right\}.
  \]
  for some $n$, $m$, $A'$ and $\vec{b}'$. Furthermore, we can assume
  that there is one column (say $j$) in $A'$ with more than one entry
  equal to $1$ (say $i$ and $i'$). Hence, $A'_{ij}$ and $A'_{i'j}$ are
  equal to $1$. Our first implementation consists of $R$ and a number
  of $c_0$ constraints,
  \[
  Q(x_{j+m}, x_i, x_{i'}) \iff R(x_1, \ldots, x_n) \bigwedge_{\substack{k: \ m+1 \leq k \leq n \\ k \not \in \{j+m, i, i'\} }} c_0(x_k) .
  \]
  This implementation leaves us with three cases, the first one is $Q
  = EQ^3$, in which case we are done. The other two cases are $Q =
  \{(0,0,1), (1,1,0)\}$ and $Q = \{(0,1,0), (1,0,1)\}$. We will give
  an implementation of $EQ^3$ with the first case, the other one is
  similar. Note that
  \[
  EQ^3(y_1, y_2, y_3) \iff \exists z,z':  Q(y_1, y_2, z) \land Q(y_3, z', z) .
  \]
  
  For the containment proof note that every relation in $\Gamma$ is
  the set of solutions to some non-coupled linear system of equations
  over GF(2). The set of feasible solutions to an instance of
  \prob{W-Max Ones$(\Gamma)$-$2$} is therefore the set of solutions to
  a linear system of equations over GF(2) with the property that every
  variable occurs at most twice. This problem is solvable by Edmonds
  and Johnson's method~\cite{well-solved}.  \qed
\end{proof}

\begin{corollary} \label{cor:il2-2}
  Let $\Gamma$ be a conservative constraint language such that
  $\Pol(\Gamma) = \Pol(IL_2)$ if there is a relation $R \in \Gamma$
  such that $R$ is not a $\Delta$-matroid relation then \prob{W-Max
    Ones$(\Gamma)$-$2$} is \cc{APX}-complete, otherwise \prob{W-Max
    Ones$(\Gamma)$-$2$} is in \cc{PO}.
\end{corollary}
\begin{proof}
  Given a constraint language $\Gamma$ such that $\Pol(\Gamma) =
  \Pol(IL_2)$ then \prob{W-Max Ones$(\Gamma)$} is
  \cc{APX}-complete~\cite{cspapprox}.

  It is not hard to see that for a relation $R \in \Gamma$, $R$ is not
  a $\Delta$-matroid relation if and only if $R$ is the set of
  solutions to a coupled system of equations. (The ``if''-part follows
  directly from the representation of relation $Q$ in
  Lemma~\ref{lem:il2-2}.)
  
  Hence, if there is a relation $R \in \Gamma$ such that $R$ is not a
  $\Delta$-matroid relation then we get \cc{APX}-completeness for
  \prob{W-Max Ones$(\Gamma)$-$2$} from Lemma~\ref{lem:il2-2}. On the
  other hand, if there is no non-$\Delta$-matroid relation in $\Gamma$
  then no relation is the set of solutions to a coupled system of
  equations and hence we get tractability from Lemma~\ref{lem:il2-2}.
  \qed
\end{proof}

The final sub-case is $\Gamma \subseteq IE_2$.

\begin{proof}[Of Lemma~\ref{lem:abc5}]
  For the containment note that the algorithm in Lemma~\ref{lem:alg}
  can be used, as an instance of \prob{W-Max Ones$(\{c_0, c_1,
    R\})$-$2$} can easily be reduced to an instance of \prob{W-Max
    Ones$(\{c_0, c_1, NAND^2\})$-$4$}.

  We will do a reduction from \prob{Max 2SAT-3} (i.e., \prob{Max 2SAT}
  where every variable occurs at most three times), which is
  \cc{APX}-complete~\cite[Chap.~8]{Kannetal99}. The reduction is based on
  Theorem~1 in~\cite{mis-3}, which in turn is based on some of Viggo
  Kann's work on 3-dimensional matching~\cite{3-dim-matching}.
  
  We will do the reduction in two steps, we will first reduce
  \prob{Max 2SAT-3}, to a restricted variant of \prob{MIS-3}. More
  precisely the graphs produced by the reduction will have maximum
  degree three and it will be possible to ``cover'' the graphs with
  $R$ (we will come back to this soon).
  
  Let $I = (V, C)$ be an arbitrary instance of \prob{Max 2SAT-3}. We
  will construct an instance $I' = (G, w)$, where $G = (V', E')$ and
  $w : V' \rightarrow \mathbb{Q}$, of weighted maximum independent
  set. We can assume, without loss of generality, that each variable
  in $I$ occurs at least once unnegated and at least once negated. For
  a node $v \in V$ construct four paths with three nodes each.
  Sequentially label the nodes in path number $x$ by $p_{x1}, p_{x2},
  p_{x3}$.  Construct three complete binary trees with four leaves
  each and label the roots of the trees with $v_1, \lnot v_1, v_2$ (or
  $v_1, \lnot v_1, \lnot v_2$ if $v$ occurs once unnegated and twice
  negated). Finally, identify the leaves of each of the trees with the
  nodes in the paths with similar labels, where two labels $p_{xy}$
  and $p_{uv}$ are similar if $y=v$.  Figure~\ref{fig:gadget} contains
  this gadget for our example variable, $v$.

\begin{figure}[hbt]
  \centering \psset{unit=0.4cm} \input{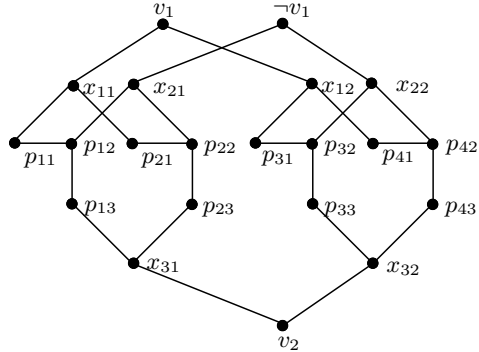}
  \caption{The graph gadget for the variable $v$ which occurs three times, two times unnegated and one time negated.}
  \label{fig:gadget}
\end{figure}
  
  Let the $w$ be defined as follows, $w(p_{12}) = w(p_{22}) = w(p_{32})
  = w(p_{42}) = 2.25$, $w(x_{21}) = w(x_{22}) = 2$ and $w(\cdot) = 1$
  otherwise.
  
  Denote the disjoint union of those paths and trees for all
  variables by $X$. A solution $S$ for the independent set problem for
  $X$ will be called \emph{consistent} if for each variable, $v$,
  (which occurs twice unnegated and once negated) we have $v_1, v_2
  \in S$ and $\lnot v_1 \not \in S$ or vice versa (i.e., $v_1, v_2
  \not \in S$ and $\lnot v_1 \in S$). It is not hard to verify (e.g.,
  with a computer assisted search) that the optimal solutions to $X$
  are consistent. Furthermore, for each consistent solution there is a
  solution which is optimal and includes or excludes the $v_i$'s and
  $\lnot v_i$'s in the same way.

  For each clause $c \in C$, containing the literals $l_1$ and $l_2$,
  add two fresh nodes $l_1$ and $l_2$ to $G'$. Connect $l_1$ and
  $l_2$ with an edge and connect $l_1$ with the node which is labelled
  with this literal (one of the roots of the trees). Do the same thing
  for $l_2$.
  
  We deduce that given a solution to $I'$ it is possible to construct
  a consistent solution with a measure which is greater than or equal to the
  measure of the original solution. The only case we have to be
  careful about is when we are given a solution $S$ where $v_1, v_2,
  \lnot v_1 \not \in S$. In this case the measure of the gadget is
  strictly less than the locally optimal solution. Hence, we can add
  $\lnot v_1$ which, in the worst case, will force us to remove one
  node which was attached to $\lnot v_1$ due to the clause which
  $\lnot v_1$ is in. However, this loss will be made up for as we can
  assign an optimal solution to the gadget.

  We have $\opt(I') \leq |V|K + \opt(I)$ where $K = 14$ is the optimum
  value for our gadget. As $\opt(I) \geq |C|/2$ and $|V| \leq 3|C|$ we
  get $\opt(I') \leq 3K|C| + \opt(I) \leq (6K+1) \opt(I)$, hence
  $\beta = 6K+1$ is an appropriate parameter for an $L$-reduction.
  
  For any consistent solution $S'$ to $I'$ we can construct a solution
  $s$ to $I$ as follows, for each variable $v \in V$ let $s(v) =
  \textsc{True}$ if $v_i \not \in S'$ and $s(v) = \textsc{False}$
  otherwise. We will then have $|\opt(I) - m(I, s)| = |\opt(I') -
  m(I', S')|$. Hence, $\gamma = 1$ is an appropriate parameter for the
  $L$-reduction.
  
  Using $c_0$ and $R$ it is possible to 2-represent $NAND^2(x, y)$. To
  reduce $I'$ to an instance of \prob{W-Max Ones$(\{R\})$-$2$} note
  that we can ``cover'' each variable gadget with $R$ and $NAND^2$,
  see Figure~\ref{fig:cover} how this is done. Furthermore, in the
  covering we have only used $v_1$, $\lnot v_1$ and $v_2$ once so it
  wont be any problems with connecting the gadgets to each other with
  $NAND^2$ constraints. \qed

\begin{figure}[hbt]
  \centering \psset{unit=0.4cm} \input{mis3-cover2.diatex}
  \begin{tabular}{l|l}
    Part of graph & Constraints \\
    \hline
    Paths & $R(p_{11}, p_{12}, p_{13}), R(p_{21}, p_{22}, p_{23})$ \\
    & $R(p_{31}, p_{32}, p_{33}), R(p_{41}, p_{42}, p_{43})$ \\
    Tree for $v_1$  & $R(p_{11}, x_{11}, p_{21}), R(p_{31}, x_{12}, p_{41}), R(x_{11}, v_1, x_{12})$ \\
    Tree for $\lnot v_1$ & $R(p_{12}, x_{21}, p_{22}), R(p_{32}, x_{22}, p_{42}), R(x_{21}, \lnot v_1, x_{22})$ \\
    Tree for $v_2$  & $R(p_{13}, x_{31}, p_{23}), R(p_{33}, x_{32}, p_{43}), R(x_{31}, v_2, x_{32})$
  \end{tabular}
  \caption{The gadget for the variable $v$ covered by the relation $R$. Note that each variable occurs
    at most twice and that $v_1$, $v_2$ and $\lnot v_1$ occurs once.
    Constraints with overlapping nodes are represented by two
    different line styles in the graph: solid and dotted.}
  \label{fig:cover}
\end{figure}  
\end{proof}

We need the following result which has been proved by
Feder~\cite[Theorem~3, fact~1]{fanout-limitations}.
\begin{lemma} \label{lem:feder-ie}
  Given a relation $R$ which is not closed under $f(x, y) = x \lor y$,
  then $R$ can $2$-represent either $NAND^2$ or $x \neq y$.
\end{lemma}

\begin{corollary} \label{cor:ie-implements-nand}
  Given a relation $R \in IE_2$ which is not closed under $f(x, y) = x
  \lor y$, then $R$ can $2$-represent $NAND^2$.
\end{corollary}
\begin{proof}
  From Lemma~\ref{lem:feder-ie} we deduce that $R$ can $2$-represent
  either $NAND^2$ or $x \neq y$, but the latter is not contained in
  $IE_2$, hence we must have the former. \qed
\end{proof}

\begin{lemma} \label{lem:delta-hard}
  Let $\Gamma$ be a conservative constraint language, if $IS_{12}^2
  \subseteq \langle \Gamma \rangle \subseteq IE_2$, and there is a
  relation $R \in \Gamma$ such that $R$ is not a $\Delta$-matroid
  relation, then \prob{W-Max Ones$(\Gamma)$-$2$} is \cc{APX}-hard.
\end{lemma}
Some parts of the following proof is similar to Feder's proof
in~\cite{fanout-limitations} that non-$\Delta$-matroids causes
\prob{Csp$(\cdot)$-$2$} to be no easier than \prob{Csp$(\cdot)$}.

\begin{proof}
  As $R$ is not a $\Delta$-matroid relation there exists tuples
  $\tup{t}, \tup{t'} \in R$ such that $d_H(\tup{t}, \tup{t'}) \geq 3$
  and a step $\tup{s} \not \in R$ from $\tup{t}$ to $\tup{t'}$ such
  that no step from $\tup{s}$ to $\tup{t'}$ is contained in $R$.
  
  Let $n$ be the arity of $R$ and let $X \subseteq [n]$ be the set of
  coordinates where $\tup{t}$ differs from $\tup{t'}$, i.e., $\tup{t}
  = \tup{t'} \oplus X$. Furthermore, let $k \in [n]$ be the coordinate
  where $\tup{s}$ differs from $\tup{t}$.
  
  By using projections and the $c_0$ and $c_1$ constraints together
  with $R$ we can $2$-represent a new relation, $P$, which is not a
  $\Delta$-matroid relation and has arity $3$. To do this, choose a
  subset $X' \subset X$ of minimal cardinality such that $k \in X'$
  and $\tup{t} \oplus X' \in R$. Note that $|X'| \geq 3$. Let $a$ and
  $b$ be two distinct coordinates in $X'$ which differs from $k$.
  Construct $P$ as follows:
  \[
  P(x_k, x_a, x_b) \iff R(x_1, x_2, \ldots, x_n)
  \bigwedge_{\substack{l \in [n] \setminus X' \\ \tup{t}[l] = 1}} c_1(x_l)
  \bigwedge_{\substack{l \in [n] \setminus X' \\ \tup{t}[l] = 0}} c_0(x_l) .
  \]
  Furthermore, let $\tup{v} = \proj{\tup{t}}{\{k,a,b\}}$ and $\tup{v'}
  = \proj{\tup{t'}}{\{k,a,b\}}$ we then have $\tup{v}, \tup{v'} \in P$
  and $\tup{v} \oplus 1, \tup{v} \oplus \{1,2\}, \tup{v} \oplus
  \{1,3\} \not \in P$. Hence, depending on $\tup{v}$ and which other
  tuples that are in $P$ we get a number of possibilities. We will use
  the following notation: $\tup{a} = \tup{v} \oplus 2$, $\tup{b} =
  \tup{v} \oplus 3$ and $\tup{c} = \tup{v} \oplus \{2,3\}$. Zero or
  more of $\tup{a}$, $\tup{b}$ and $\tup{c}$ may be contained in $P$.
  Tables~\ref{tab:delta-hard-start}--\ref{tab:delta-hard-end} list the
  possible relations we can get, up to permutations of the
  coordinates. Note that $\tup{a} \in P, \tup{b}, \tup{c} \not \in P$
  and $\tup{b} \in P, \tup{a}, \tup{c} \not \in P$ are equivalent if
  we disregard permutations of the coordinates. Similarly $\tup{a},
  \tup{c} \in P, \tup{b} \not \in P$ and $\tup{b}, \tup{c} \in P,
  \tup{a} \not \in P$ are equivalent.
  
  Some of the relations are not in $IE_2$ and can therefore be omitted
  from further consideration (it is clear that if $P$ is not in $IE_2$
  then $R$ is not in $IE_2$ either, which is a contradiction with the
  assumptions in the lemma). Others can $2$-represent $EQ^3$, or can
  do so together with $NAND^2$. As an example consider $A5$, then
  \begin{align}
    \exists y_1,y_2,y_3,z_1,z_2,z_3: &A5(y_1, x_1, z_1) \land NAND^2(z_1, y_2) \land \notag \\
                                     &A5(y_2, x_2, z_2) \land NAND^2(z_2, y_3) \land \notag \\
                                     &A5(y_3, x_3, z_3) \land NAND^2(z_3, y_1)       \notag
  \end{align}
  is a $2$-representation of $EQ^3(x_1, x_2, x_3)$. Similar
  constructions works for some of the other relations. If we can
  $2$-represent $EQ^3$ then we get \cc{poly-APX}-hardness due to the
  construction in Lemma~\ref{lem:is-implements-nand},
  Lemma~\ref{lem:k-repr} and a simple reduction from \prob{MIS}.
  Information about which relations this applies to is contained in
  Table~\ref{tab:non-d-sum}.
  
  Furthermore, some of the relations can $2$-represent other relations
  in the table, see Table~\ref{tab:non-delta-impl} for those. This
  implies that the only relation that is left to prove
  \cc{APX}-hardness for is $ABC1$. We will do this with a reduction
  from \prob{MIS-3}. Let $G = (V, E)$ be an instance of \prob{MIS-3},
  we will construct an instance $I' = (V', C', w')$ of \prob{W-Max
    Ones$(\Gamma)$-$2$} with the assumption that $ABC1 \in \Gamma$.
  Furthermore, due to Lemma~\ref{lem:is-implements-nand} and
  Corollary~\ref{cor:ie-implements-nand} we are free to assume that
  $NAND^2 \in \Gamma$.  For every variable $v \in V$, if there are
  three occurrences of $v$ in $I$ add one fresh variable for each
  occurrence of $v$ in $I$ to $V'$, name those fresh variables $v_1,
  v_2$ and $v_3$. If there are less than three occurrences add $v$ to
  $V'$. Furthermore, for each edge $(v, x)$ for some $x \in V$ add a
  $NAND^2(v_i, x_j)$ constraint to $C'$. So far $I'$ is an instance
  where each variable occurs at most twice and the variables which
  corresponds to nodes in $G$ with degree three occurs once in $I'$.
  
  For each node $v \in V$ with degree three add the constraint
  $ABC1(v_1, v_2, v_3)$ to $C'$. Finally, let $w(x) = 1$ for every $x
  \in V$ with degree less than three and $w(v_1) = 1$ and $w(v_2) =
  w(v_3) = 0$ for every $v \in V$ with degree three. For every
  solution $s$ to $I'$ we can construct a solution $S$ to $I$ such
  that $m(I', s) = m(I, S)$ to see this note that if $s(x) = 1$ for
  some variable $x$ then due to the $ABC1$ constraints the other
  occurrences of $x$ also have the value $1$.  On the other hand, if
  $s(x) = 0$ then we can set the other occurrences of $x$ to $0$
  without changing the measure of the solution and without conflicts
  with any constraints. This implies that there is an $S$-reduction
  from \prob{MIS-3} to \prob{W-Max Ones$(\Gamma)$-$2$}. \qed
\end{proof}
    The results obtained in Lemma~\ref{lem:delta-hard} is not optimal
    for all non-$\Delta$-matroids. It is noted in the proof that we
    get \cc{poly-APX}-hardness results for some of the relations, but
    we do not get this for all of them. In particular we do not get
    this for $A3$, $AB1$, $BC4$, $ABC1$, $ABC3$, $ABC5$ and $ABC6$.
    However, $ABC5$ is contained in \cc{APX} by Lemma~\ref{lem:abc5}.

\begin{table}
  \begin{tabular}{ll||ll}
    Relation & Implementation or & Relation & Implementation or \\
             & comment           &          & comment           \\
    \hline
    \hline
    1        &                 $EQ^2$     & BC2      & Not in $IE_2$                            \\
    2        & Not in $IE_2$              & BC3      & Not in $IE_2$                            \\
    \cline{1-2} A1&            $EQ^2$     & BC4      & See Table~\ref{tab:non-delta-impl}       \\
    \cline{3-4} A2& Not in $IE_2$         & AB1      & See Table~\ref{tab:non-delta-impl}       \\
    A3       & See Table~\ref{tab:non-delta-impl} & AB2 & Not in $IE_2$                         \\
    A4       & Not in $IE_2$              & AB3      & Not in $IE_2$                            \\
    A5       &                 $NAND^2$   & AB4      & Not in $IE_2$                            \\
    A6       &                 $EQ^2$     & AB5      & Not in $IE_2$                            \\
    \cline{1-2} C1&            $EQ^2$     & AB6      & Not in $IE_2$                            \\
    \cline{3-4} C2&            $NAND^2$   & ABC1     & See Lemma~\ref{lem:delta-hard}           \\
    C3       & Not in $IE_2$              & ABC2     & Not in $IE_2$                            \\
    C4       & Not in $IE_2$              & ABC3     & See Table~\ref{tab:non-delta-impl}       \\
    C5       & Not in $IE_2$              & ABC4     & Not in $IE_2$                            \\
    C6       &                 $EQ^2$     & ABC5     & See Lemma~\ref{lem:abc5}                 \\
    \cline{1-2} BC1&           $EQ^2$     & ABC6     & See Table~\ref{tab:non-delta-impl}                        
  \end{tabular}
  \caption{Non-$\Delta$-matroid relations in Lemma~\ref{lem:delta-hard}. If there is a relation in the
    ``Implementation or comment'' column then this relation can $2$-represent $EQ^3$ together with the
    noted relation. If this second relation is $EQ^2$ then the relation can in fact $2$-represent $EQ^3$ on
    its own, $EQ^2$ is not needed. \label{tab:non-d-sum}}
\end{table}
\begin{table}
  \begin{tabular}{lll}
    Relation & Implements & Implementation \\
    \hline
    A3       & $ABC5$& $\exists x': A3(x_1, x', x_3) \land NAND^2(x', x_2)$   \\
    AB1      & $ABC5$& $\exists x',x'': AB1(x_1, x', x'') \land NAND^2(x', x_2) \land NAND^2(x'', x_3)$ \\
    BC4      & $ABC5$& $\exists x': BC4(x_1, x', x_2) \land NAND^2(x', x_3)$  \\
    ABC3     & $ABC5$& $\exists x': ABC3(x_1, x', x_3) \land NAND^2(x', x_2)$ \\
    ABC6     & $ABC1$& $\exists x': ABC6(x', x_2, x_3) \land NAND^2(x', x_1)$
  \end{tabular}
  \caption{Implementations in Lemma~\ref{lem:delta-hard} \label{tab:non-delta-impl}} 
\end{table}
  \begin{table}
    \begin{minipage}[b]{0.1\linewidth}%
      \centering%
      \begin{tabular}{l}
        111 \\
        000 \\
        \hline
      \end{tabular}
      \\1
    \end{minipage}%
    \begin{minipage}[b]{0.1\linewidth}%
      \centering%
      \begin{tabular}{l}
        110 \\
        001 \\
        \hline
      \end{tabular}
      \\2
    \end{minipage}
    %
    \begin{minipage}[b]{0.1\linewidth}%
      \centering%
      \begin{tabular}{l}
        000 \\
        111 \\
        010 \\
        \hline
      \end{tabular}
      \\A1
    \end{minipage}
    \begin{minipage}[b]{0.1\linewidth}%
      \centering%
      \begin{tabular}{l}
        100 \\
        011 \\
        110 \\
        \hline
      \end{tabular}
      \\A2
    \end{minipage}
    \begin{minipage}[b]{0.1\linewidth}%
      \centering%
      \begin{tabular}{l}
        010 \\
        101 \\
        000 \\
        \hline
      \end{tabular}
      \\A3
    \end{minipage}
    \begin{minipage}[b]{0.1\linewidth}%
      \centering%
      \begin{tabular}{l}
        110 \\
        001 \\
        100 \\
        \hline
      \end{tabular}
      \\A4
    \end{minipage}
    \begin{minipage}[b]{0.1\linewidth}%
      \centering%
      \begin{tabular}{l}
        101 \\
        010 \\
        111 \\
        \hline
      \end{tabular}
      \\A5
    \end{minipage}
    \begin{minipage}[b]{0.1\linewidth}%
      \centering%
      \begin{tabular}{l}
        111 \\
        000 \\
        101 \\
        \hline
      \end{tabular}
      \\A6
    \end{minipage}
    \caption{Non-$\Delta$-matroid relations where $\tup{a}, \tup{b}, \tup{c} \not \in P$
             followed by the relations where $\tup{a} \in P$ and $\tup{b}, \tup{c} \not \in P$.  \label{tab:delta-hard-start}}
  \end{table}
  \begin{table}
    \begin{minipage}[b]{0.1\linewidth}%
      \centering%
      \begin{tabular}{l}
        000 \\
        111 \\
        011 \\
        \hline
      \end{tabular}
      \\C1
    \end{minipage}%
    \begin{minipage}[b]{0.1\linewidth}%
      \centering%
      \begin{tabular}{l}
        100 \\
        011 \\
        111 \\
        \hline
      \end{tabular}
      \\C2
    \end{minipage}%
    \begin{minipage}[b]{0.1\linewidth}%
      \centering%
      \begin{tabular}{l}
        010 \\
        101 \\
        001 \\
        \hline
      \end{tabular}
      \\C3
    \end{minipage}%
    \begin{minipage}[b]{0.1\linewidth}%
      \centering%
      \begin{tabular}{l}
        110 \\
        001 \\
        101 \\
        \hline
      \end{tabular}
      \\C4
    \end{minipage}%
    \begin{minipage}[b]{0.1\linewidth}%
      \centering%
      \begin{tabular}{l}
        011 \\
        100 \\
        011 \\
        \hline
      \end{tabular}
      \\C5
    \end{minipage}%
    \begin{minipage}[b]{0.1\linewidth}%
      \centering%
      \begin{tabular}{l}
        111 \\
        000 \\
        100 \\
        \hline
      \end{tabular}
      \\C6
    \end{minipage}%
    \caption{Non-$\Delta$-matroid relations where only $\tup{c} \in P$.}
  \end{table}
  \begin{table}
    \begin{minipage}[b]{0.1\linewidth}%
      \centering%
      \begin{tabular}{l}
        000 \\
        111 \\
        001 \\
        011 \\
        \hline
      \end{tabular}
      BC1
    \end{minipage}%
    \begin{minipage}[b]{0.1\linewidth}%
      \centering%
      \begin{tabular}{l}
        100 \\
        011 \\
        101 \\
        111 \\
        \hline
      \end{tabular}
      BC2
    \end{minipage}%
    \begin{minipage}[b]{0.1\linewidth}%
      \centering%
      \begin{tabular}{l}
        010 \\
        101 \\
        011 \\
        001 \\
        \hline
      \end{tabular}
      BC3
    \end{minipage}%
    \begin{minipage}[b]{0.1\linewidth}%
      \centering%
      \begin{tabular}{l}
        001 \\
        110 \\
        000 \\
        010 \\
        \hline
      \end{tabular}
      BC4
    \end{minipage}
%
    \begin{minipage}[b]{0.1\linewidth}%
      \centering%
      \begin{tabular}{l}
        000 \\
        111 \\
        010 \\
        001 \\
        \hline
      \end{tabular}
      AB1
    \end{minipage}%
    \begin{minipage}[b]{0.1\linewidth}%
      \centering%
      \begin{tabular}{l}
        100 \\
        011 \\
        110 \\
        101 \\
        \hline
      \end{tabular}
      AB2
    \end{minipage}%
    \begin{minipage}[b]{0.1\linewidth}%
      \centering%
      \begin{tabular}{l}
        010 \\
        101 \\
        000 \\
        011 \\
        \hline
      \end{tabular}
      AB3
    \end{minipage}%
    \begin{minipage}[b]{0.1\linewidth}%
      \centering%
      \begin{tabular}{l}
        110 \\
        001 \\
        100 \\
        111 \\
        \hline
      \end{tabular}
      AB4
    \end{minipage}%
    \begin{minipage}[b]{0.1\linewidth}%
      \centering%
      \begin{tabular}{l}
        011 \\
        100 \\
        001 \\
        010 \\
        \hline
      \end{tabular}
      AB5
    \end{minipage}%
    \begin{minipage}[b]{0.1\linewidth}%
      \centering%
      \begin{tabular}{l}
        111 \\
        000 \\
        101 \\
        110 \\
        \hline
      \end{tabular}
      AB6
    \end{minipage}%
    \caption{Non-$\Delta$-matroid relations where $\tup{b}, \tup{c} \in P$ and
             $\tup{a} \not \in P$ followed by relations where $\tup{a}, \tup{b} \in P$ and $\tup{c} \not \in P$.}
  \end{table}
  \begin{table}
    \begin{minipage}[b]{0.1\linewidth}%
      \centering%
      \begin{tabular}{l}
        000 \\
        111 \\
        010 \\
        001 \\
        011 \\
        \hline
      \end{tabular}
      ABC1
    \end{minipage}
    \begin{minipage}[b]{0.1\linewidth}%
      \centering%
      \begin{tabular}{l}
        100 \\
        011 \\
        110 \\
        101 \\
        111 \\
        \hline
      \end{tabular}
      ABC2
    \end{minipage}
    \begin{minipage}[b]{0.1\linewidth}%
      \centering%
      \begin{tabular}{l}
        010 \\
        101 \\
        000 \\
        011 \\
        001 \\
        \hline
      \end{tabular}
      ABC3
    \end{minipage}
    \begin{minipage}[b]{0.1\linewidth}%
      \centering%
      \begin{tabular}{l}
        110 \\
        001 \\
        100 \\
        111 \\
        101 \\
        \hline
      \end{tabular}
      ABC4
    \end{minipage}
    \begin{minipage}[b]{0.1\linewidth}%
      \centering%
      \begin{tabular}{l}
        011 \\
        100 \\
        001 \\
        010 \\
        000 \\
        \hline
      \end{tabular}
      ABC5
    \end{minipage}
    \begin{minipage}[b]{0.1\linewidth}%
      \centering%
      \begin{tabular}{l}
        111 \\
        000 \\
        101 \\
        110 \\
        100 \\
        \hline
      \end{tabular}
      ABC6
    \end{minipage}
    \caption{Non-$\Delta$-matroid relations where $\tup{a}, \tup{b}, \tup{c} \in P$.\label{tab:delta-hard-end}}
    \end{table}

We are now finally ready to state the proof of the classification
theorem for two variable occurrences.

\begin{proof}[Of Theorem~\ref{th:classification-2}, part 1]
Follows from Khanna et al's results on \prob{W-Max Ones}~\cite{cspapprox}. \qed
\end{proof}
\begin{proof}[Of Theorem~\ref{th:classification-2}, part 2]
  Follows from Corollary~\ref{cor:il2-2}, Lemma~\ref{lem:id2-hard} and
  Lemma~\ref{lem:id2-po}. \qed
\end{proof}
\begin{proof}[Of Theorem~\ref{th:classification-2}, part 3]
  Follows from Lemma~\ref{lem:delta-hard}. \qed
\end{proof}
\begin{proof}[Of Theorem~\ref{th:classification-2}, part 3]
  Follows from~\cite[Theorem~4]{fanout-limitations}. \qed
\end{proof}

\section*{Proofs for Results in Section~\ref{sec:non-cons}}
\begin{proof}[Of Theorem~\ref{th:non-cons}]
  Let $\Gamma$ be a non-1-valid constraint language and $k$ an integer
  such that \prob{W-Max Ones$(\Gamma \cup \{c_0, c_1\})$-$k$} (this
  problem will hereafter be denoted by $\Pi_{01}$) is \cc{NP}-hard. We
  will prove the theorem with a reduction from $\Pi_{01}$ to \prob{W-Max
    Ones$(\Gamma)$-$k$} (hereafter denoted by $\Pi$).

  As $\Gamma$ is not 1-valid there exists a relation $R \in \Gamma$
  such that $(1, \ldots, 1) \not \in R$. Let $r$ be the arity of $R$
  and let $\tup{t}$ be the tuple in $R$ with the maximum number of
  ones.  Assume, without loss of generality, that $\tup{t} = (0, 1,
  \ldots, 1)$.
  
  The assumption in the theorem implies that it is \cc{NP}-hard to
  decide the following question: given an instance $I = (V, C, w)$ of
  $\Pi_{01}$ and an integer $K$ is $\opt(I) \geq K$?
  
  Let $I = (V, C, w), K$ be an arbitrary instance of the decision
  variant of $\Pi_{01}$. We will transform $I$ into an instance $I' =
  (V', C', w'), K'$ of the decision variant of $\Pi$ by first removing
  constraint applications using $c_0$ and then removing constraint
  applications using $c_1$.
  
  At the start of the reduction let $V' = V$ and $C' = C$. For each
  constraint $(c_0, (v)) \in C'$ replace this constraint with $(R, (v,
  v_1, \ldots, v_{r-1}))$ where $v_1, \ldots, v_{r-1}$ are fresh
  variables, furthermore add the constraint $(c_1, (v_k))$ for $k = 1,
  \ldots, r-1$ to $C'$.
  
  Let $c$ be the number of variables which are involved in $c_1$
  constraints. For each constraint using $c_1$, $(c_1, (v)) \in C'$,
  remove this constraint and set $w'(v) = L + w(v)$, where $L$ is a
  sufficiently large integer ($L = 1 + \sum_{v \in V} w(v)$ is enough).
  For every variable $v$ which is not involved in a $c_1$ constraint
  let $w'(v) = w(v)$.
  
  Finally let $K' = K + cK$. Given a solution $s'$ to $I'$ such that
  $m(I', s') \geq K'$ it is clear that this solution also is a
  solution to $I$ such that $m(I, s') \geq K$. Furthermore, if there
  is a solution $s$ to $I$ such that $m(I, s) \geq K$ then $s$ is a
  solution $I'$ such that $m(I', s) \geq K'$.  \qed
\end{proof}

\bibliography{bibtexdb}

\end{document}